\documentclass[epj,amsfonts,amsmath,twocolumn,superscriptaddress,nofootinbib]{revtex4}

\usepackage{amssymb,color,paralist,amsmath,epsfig}
\usepackage{graphicx}
\usepackage{comment}
\usepackage{amsfonts}
\usepackage{relsize}
\usepackage{nicefrac,esint}
\usepackage{float}

\usepackage[colorlinks,citecolor=blue,linktoc=all,linkcolor=cyan]{hyperref}

\newcommand{\hp}{{+}}
\newcommand{\hm}{{-}}

\newcommand{\bk}{{\boldsymbol k}}

\newcommand{\cG}{{\cal G}}
\newcommand{\cF}{{\cal F}}
\newcommand{\cma}{{\theta_{\mathrm{cm}}}}

\newcommand{\beq}{\begin{equation}}
\newcommand{\eeq}{\end{equation}}

\newcommand{\ber}{\begin{eqnarray}}
\newcommand{\eer}{\end{eqnarray}}

\begin{document}

\title{Dispersion relation formalism for the two-photon exchange correction to elastic muon-proton scattering: Elastic intermediate state}

\author{Oleksandr Tomalak}
\affiliation{Institut f\"ur Kernphysik \& Cluster of Excellence PRISMA, Johannes Gutenberg Universit\"at Mainz, 55128 Mainz, Germany}
%\author{Barbara Pasquini}
%\affiliation{Dipartimento di Fisica, Universit\`a degli Studi di Pavia, Pavia, Italy}
%\affiliation{INFN Sezione di Pavia, Pavia, Italy}
%\author{Marc Vanderhaeghen}
%\affiliation{Institut f\"ur Kernphysik and PRISMA Cluster of Excellence, Johannes Gutenberg Universit\"at, Mainz, Germany}
%\affiliation{Institut f\"ur Kernphysik \& Cluster of Excellence PRISMA, Johannes Gutenberg Universit\"at Mainz, 55128 Mainz, Germany}
%\affiliation{PRISMA Cluster of Excellence, Johannes Gutenberg-Universit\"at,  Mainz, Germany}
%\affiliation{Department of Physics, Taras Shevchenko National University of Kyiv, Ukraine}

\author{Marc Vanderhaeghen}
\affiliation{Institut f\"ur Kernphysik \& Cluster of Excellence PRISMA, Johannes Gutenberg Universit\"at Mainz, 55128 Mainz, Germany}
%\affiliation{PRISMA Cluster of Excellence, Johannes Gutenberg-Universit\"at,  Mainz, Germany}

\date{\today}

\begin{abstract}
We evaluate the two-photon exchange correction to the unpolarized cross section in the elastic muon-proton scattering within dispersion relations. One of the six independent invariant amplitudes requires a subtraction. We fix the subtraction function to the model estimate of the full two-photon exchange at one of three MUSE beam energies and make a prediction for the two other energies. Additionally, we present single and double polarization observables accounting for the lepton mass.
\end{abstract}

\maketitle

\tableofcontents

\section{Introduction}
\label{sec1}

The forthcoming muon-proton scattering experiment (MUSE) \cite{Gilman:2013eiv,Gilman:2017hdr} aims to shed a new light on the "proton radius puzzle", the discrepancy in the extracted proton charge radius from the hydrogen spectroscopy \cite{Mohr:2012tt} and electron-proton scattering \cite{Bernauer:2010wm, Bernauer:2013tpr} versus extractions from the Lamb shift in muonic hydrogen \cite{Pohl:2010zza, Antognini:1900ns}. MUSE is going to complement this picture by providing the first measurement of the charge radius from the elastic muon-proton scattering. \footnote{Note that the COMPASS collaboration is planning to probe the elastic muon-proton scattering at low momentum transfer and high energy  \cite{COMPASS:2017}. Alternatively, the muon-proton interaction is going to be tested in measurements of the ratio of the muon- and electron-pair production to electron-pair production cross sections at MAMI \cite{Pauk:2015oaa,Heller:2018ypa}.}

MUSE will scatter electrons, positrons, muons and antimuons on the proton target and aims to determine cross sections, two-photon effects, form factors, and radii independently in $e p$ and $\mu p$ scattering \cite{Gilman:2017hdr}. To achieve the required sub-percent accuracy, all radiative corrections at the 1-loop level, at least, should be carefully accounted for.

The standard electron-proton scattering QED 1-loop radiative corrections are described and collected in Refs. \cite{Tsai:1961zz,Maximon:2000hm,Vanderhaeghen:2000ws,Gramolin:2014pva}. The numerical estimate of QED radiative corrections in the soft-photon approximation was recently performed in Ref. \cite{Koshchii:2017dzr}. In Ref. \cite{Talukdar:2017dhw}, it was shown that the commonly used {\it peaking approximation} for the lepton-proton bremsstrahlung is not applicable for muon-proton scattering at low energies of MUSE. Besides exactly calculable QED corrections, the precision of modern experiments requires an accurate knowledge of the contribution from graphs with two exchanged photons (TPE) between the lepton and proton lines beyond the approximation where one of photons is soft, which is an active research field over the last decades \cite{Guichon:2003qm,Blunden:2003sp,Gorchtein:2004ac,Pasquini:2004pv,Chen:2004tw, Afanasev:2005mp,Gorchtein:2006mq,Borisyuk:2008db,Kivel:2009eg,Kivel:2012vs,Hill:2012rh,Gorchtein:2014hla,Tomalak:2014sva,Tomalak:2015aoa,Tomalak:2015hva,Dye:2016uep,Tomalak:2016vbf,Blunden:2017nby,Tomalak:2017shs,Jones:1999rz, Gayou:2001qd,Wells:2000rx,Maas:2004pd,Punjabi:2005wq,Puckett:2010ac,Meziane:2010xc,Guttmann:2010au,BalaguerRios:2012uk,Abrahamyan:2012cg,Waidyawansa:2013yva,Kumar:2013yoa,Nuruzzaman:2015vba,Zhang:2015kna,Rachek:2014fam,Rimal:2016toz,Henderson:2016dea}. MUSE is also going to test TPE effects at the sub-percent level by measuring scattering of particles and antiparticles.

The leading proton intermediate state TPE contribution was estimated within the hadronic model \cite{Blunden:2003sp} in the kinematics of the MUSE experiment in Ref. \cite{Tomalak:2014dja}. The contribution from all inelastic excitations in the {\it near-forward} approximation was found \cite{Tomalak:2015hva} to be an order of magnitude smaller than the elastic contribution, which is expected at energies of MUSE below the pion-production threshold. Subsequent evaluations of the $\sigma$-meson exchange correction \cite{Koshchii:2016muj,Zhou:2016psq} as well as of the $\Delta$-resonance TPE contribution \cite{Zhou:2016psq} within the hadronic model of Refs. \cite{Kondratyuk:2005kk,Kondratyuk:2007hc,Graczyk:2013pca,Zhou:2014xka,Lorenz:2014yda} confirmed the dominance of the elastic channel. 

As it was shown in elastic electron-proton scattering \cite{Borisyuk:2008db,Tomalak:2014sva,Blunden:2017nby,Tomalak:2017shs}, hadronic model calculations \cite{Blunden:2003sp,Kondratyuk:2005kk,Kondratyuk:2007hc,Graczyk:2013pca,Zhou:2014xka,Lorenz:2014yda} violate the unitarity in case of derivative photon-hadron couplings and can lead to an unphysical high-energy behavior of TPE corrections. Consequently, the dispersion relation approach \cite{Gorchtein:2006mq,Borisyuk:2008es,Borisyuk:2012he,Borisyuk:2013hja,Borisyuk:2015xma,Tomalak:2014sva,Tomalak:2016vbf,Blunden:2017nby,Tomalak:2017shs} is a favourable way to treat TPE corrections as a sum of different intermediate channels.

In this work, we introduce the dispersion relation framework to evaluate TPE corrections in the elastic muon-proton scattering. To write down dispersion relations, we study unitarity constraints on the high-energy behavior of TPE amplitudes. In particular, the helicity-flip amplitude $\cF_4$, which is suppressed by the lepton mass and is therefore irrelevant for electron-proton scattering observables, does not vanish at infinite energy. Consequently, we need to subtract the dispersion relation for $\cF_4$, which is the only amplitude affected by the subtraction function in the forward doubly virtual Compton scattering \cite{Tomalak:2015aoa}. Moreover, a model estimate of this amplitude within unsubtracted dispersion relations does not satisfy the low-$Q^2$ limit of TPE contributions. As a first step in our subtracted DR framework, we account for the elastic intermediate state TPE and fix the subtraction function to the evaluation of the total TPE correction in the near-forward approximation of Ref. \cite{Tomalak:2015hva}.

The paper is organized as follows: We describe kinematics and observables in the elastic lepton-proton scattering and discuss TPE corrections in Section \ref{sec2}. In Section \ref{sec3}, we present a dispersion relation formalism to evaluate the real parts of TPE amplitudes for the case of massive lepton-proton scattering. The imaginary parts of TPE amplitudes are calculated by unitarity relations in Section \ref{sec:unitarity}. Real parts of four among six independent invariant amplitudes are reconstructed within unsubtracted dispersion relations in Section \ref{sec:dispersion}. The dispersion relation prediction for the cross-section correction requires one subtraction function. We describe how to fix it to the known TPE correction at some lepton energy in Section \ref{sec:subtracted}. We present results of the subtracted dispersion relation analysis taking the subtraction function from Ref. \cite{Tomalak:2015hva} in Section \ref{sec5}. We give our conclusions and outlook in Section \ref{sec6}. The photon-polarization density matrix is described in Appendix \ref{app:photon_polarization}. The derivation of forward and high-energy limits for TPE amplitudes is described in Appendices \ref{app:forward_relations} and \ref{app:high_energy_relations} respectively. A detailed comparison of dispersion relations to the hadronic model calculation for the proton intermediate state TPE contribution is given in Appendix \ref{app:hm_vs_drs}.

\section{Elastic muon-proton scattering and two-photon exchange}
\label{sec2}

In this Section, we describe the elastic lepton-proton scattering and two-photon exchange corrections to this process. We first discuss the kinematics with an emphasis
on the forthcoming MUSE experiment. Afterward, we present the formalism of invariant amplitudes in the assumption of discrete symmetries of QED and QCD and discuss their general properties. We provide compact expressions for the unpolarized cross section and polarization transfer observables in the one-photon exchange approximation and for the leading two-photon exchange contributions to them.

\vspace{-0.3cm}
\subsection{Kinematics in elastic muon-proton scattering}
\label{sec:kinematics_massive_lepton}
\vspace{-0.2cm}

\begin{figure}[H]
\begin{center}%\centering
\includegraphics[width=.35\textwidth]{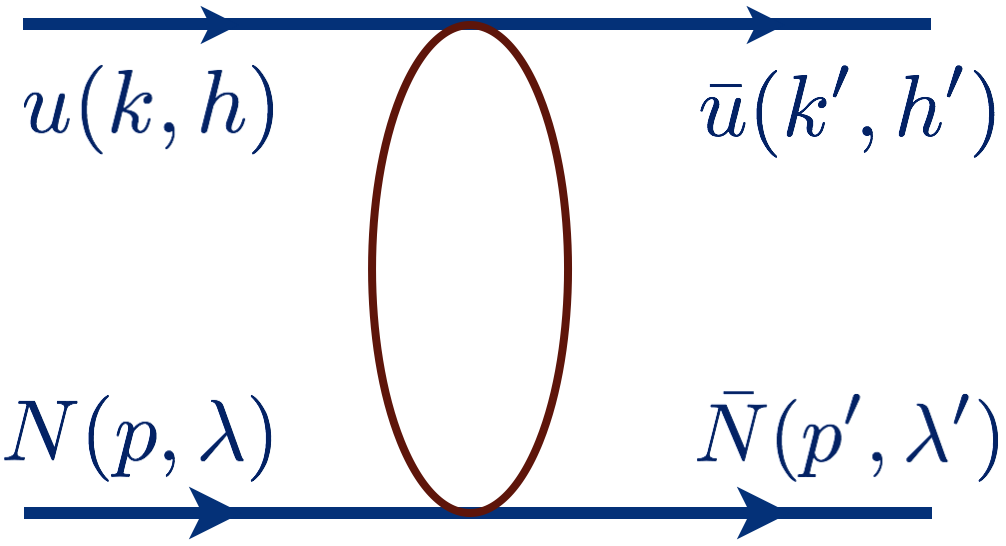}
\end{center}
\caption{Elastic lepton-proton scattering.}
\label{elastic_scattering_general}
\end{figure}

Elastic muon-proton scattering $ \mu( k , h ) + p( p, \lambda ) \to \mu( k^\prime, h^\prime) + p(p^\prime, \lambda^\prime) $,  where $ h(h^\prime) $ denote the incoming (outgoing) muon helicities and $ \lambda(\lambda^\prime) $ the corresponding proton helicities respectively (see Fig. \ref{elastic_scattering_general}), is completely described by 2 Mandelstam variables, e.g., $ Q^2 = - (k-k^\prime)^2 $ - the squared momentum transfer, and $ s = ( p + k )^2 $ - the squared energy in the lepton-proton center-of-mass (c.m.) reference frame.

The squared momentum transfer is expressed in terms of the lepton scattering angle $ \cma $ in the c.m. reference frame by
\ber
Q^2 = - ( k - k^\prime )^2 = \frac{ \Sigma_s }{2 s} ( 1 - \cos \cma )  ,
\eer
with the kinematical triangle function $\Sigma_s$:
\ber
 \Sigma_s \equiv \Sigma (s, M^2, m^2) = (s-(M +m)^2)(s-(M -m)^2),
\eer
where $ M (m)$ denotes the proton (muon) mass respectively.

In terms of the laboratory frame momenta $ p = (M, 0), ~ k = (\omega, \bk), ~ k^\prime = (\omega^\prime, \bk^\prime), ~ p^\prime = (E_p^\prime, \bk-\bk^\prime)$, the invariant variables are expressed as 
\ber
Q^2 & = & 2 M ( \omega - \omega^\prime) , \\
s & = & M^2 + 2 M \omega + m^2,  \\
\Sigma_s & = & 4 M^2 {\bf{k}}^2.
\eer
The momentum transfer can be also determined from the laboratory frame scattering angle $ \theta_{\mathrm{lab}} $ as
\ber
Q^2 & = & \frac{ M + \omega \sin^2 \theta_{\mathrm{lab}} - \sqrt{ M^2 - m^2 \sin^2 \theta_{\mathrm{lab}} } \cos \theta_{\mathrm{lab}}}{ ( \omega + M )^2 -{\bf{k}}^2 \cos^2 \theta_{\mathrm{lab}} } \nonumber \\
&\times&  2 M {\bf{k}}^2,
\eer
with the relation between the final lepton energy $\omega'$ and scattering angle $ \theta_{\mathrm{lab}} $:
\ber
\cos \theta_{\mathrm{lab}} & = & \frac{ \omega \omega^\prime - m^2 - M (\omega - \omega^\prime)}{|\bf{k}|  |\bf{k}^\prime| }.
\eer

The kinematically allowed momentum transfer region is defined by
\ber
 0 < Q^2 < \frac{\Sigma_s}{s}.
\eer

In theoretical applications, it is convenient to introduce the crossing-symmetric variable $\nu $:
\ber
\nu = \frac{s-u}{4} = (K \cdot P) = M \frac{ \omega + \omega^\prime}{2} = M \omega - \frac{Q^2}{4},
\eer
with the $u$-channel squared energy $ u = ( k - p^\prime )^2 $ and the averaged momentum variables:
\ber
 P = \frac{p+p^\prime}{2}, ~~~ K = \frac{k+k^\prime}{2}.
 \eer
The crossing-symmetric variable $ \nu$ changes sign with $ s\leftrightarrow u $ channel crossing. 

In experiment, instead of the Mandelstam invariant $ s $ or the crossing symmetric variable $ \nu $, one can use the virtual photon polarization parameter $ \varepsilon $. Keeping the physical meaning of $ \varepsilon / \tau_P $ as a relative flux of virtual photons with longitudinal polarization in case of the one-photon exchange in any frame with collinear initial and final proton momenta, e.g., the laboratory or c.m. frame,  we express it in terms of invariants as
\ber \label{epsilon_def}
\varepsilon = \frac{\nu^2 - M^4 \tau_P ( 1 + \tau_P )}{ \nu^2 + M^4 \tau_P ( 1 + \tau_P )  ( 1 - 2 \varepsilon_0 )},
\eer 
with $\tau_P = Q^2 / (4 M^2)$ and $ \varepsilon_0 = 2m^2/Q^2 $, which can equivalently be expressed as $\varepsilon_0 = 1/(2\tau_l)$ with $\tau_l = Q^2 / (4 m^2)$. 
We discuss details of the photon-polarization density matrix in Appendix \ref{app:photon_polarization}. The photon polarization parameter $ \varepsilon $ varies between $ \varepsilon_0 < 1 $ and $1 $ for the fixed momentum transfer $ Q^2 > 2 m^2 $ and between $ 1 $ and $ \varepsilon_0  > 1 $ for the fixed momentum transfer $ Q^2 < 2 m^2 $. The high-energy limit corresponds to $ \varepsilon = 1 $. The value of the critical momentum transfer $ Q^2 = 2 m^2 $, corresponding with $ \varepsilon = 1 $ for all possible beam energies, is given by $ Q^2 \simeq 0.022 ~\mathrm{GeV}^2 $ for muon beams. This value is inside the MUSE kinematical region for all three nominal beam momenta.

The introduced parameter $\varepsilon$ differs from the degree of linear polarization of transverse photons $ \varepsilon_\mathrm{T}$:
\ber \label{transverse_epsilon}
\varepsilon_\mathrm{T}= \frac{\nu^2  - M^4 \tau_P ( 1 + \tau_P ) ( 1 + 2 \varepsilon_0 )}{ \nu^2 + M^4 \tau_P ( 1 + \tau_P )  ( 1 - 2 \varepsilon_0 )},
\eer
with $ 0 \le \varepsilon_\mathrm{T} < 1$, where $ \varepsilon_\mathrm{T} = 0 $  corresponds to the backward kinematics and $ \varepsilon_\mathrm{T} = 1$ describes the forward scattering. The difference between the two polarization parameters is suppressed by the lepton mass:
\ber
\varepsilon - \varepsilon_\mathrm{T}= \varepsilon_0 \left( 1 - \varepsilon_\mathrm{T} \right).
\eer

\vspace{0.2cm}

\subsection{Helicity amplitudes formalism}
\label{sec:HA_and_generalized_FFs}
\vspace{-0.35cm}

For the $l^- p \to l^- p $ process, there are 16 possible helicity amplitudes $ T_{h^\prime \lambda^\prime, h \lambda} $ with positive or negative helicities $h,h^\prime,\lambda,\lambda^\prime$ = $\pm $, see Fig. \ref{elastic_scattering_general}. We work with helicity amplitudes in the c.m. reference frame. The discrete symmetries of QCD and QED, i.e., parity and time-reversal invariance, leave just six independent amplitudes:
\ber
T_1 \equiv T_{\hp \hp, \hp \hp}, \quad T_2 \equiv T_{\hp \hm, \hp \hp},  \quad T_3 \equiv T_{\hp \hm, \hp \hm} , \nonumber \\ ~T_4 \equiv T_{\hm \hp, \hp \hp}, \quad T_5 \equiv T_{\hm \hm, \hp \hp},  \quad T_6 \equiv T_{\hm \hp, \hp \hm}.
\eer
Consequently, the $ l^{-} p $ elastic scattering is completely described by six generalized form factors (or invariant amplitudes) that are complex functions of two independent kinematical variables. The lepton massless limit is described by a part without the flip of lepton helicity $ T_{h^\prime \lambda^\prime, h \lambda}^{\mathrm{non-flip}} $ \cite{Guichon:2003qm}. To describe the muon-proton scattering, we have to add the part with lepton helicity flip $ T_{h^\prime \lambda^\prime, h \lambda}^{\mathrm{flip}} $, which is proportional to the mass of the lepton \cite{Goldberger:1957ac,Gorchtein:2004ac}. The resulting amplitude is given by the sum of these two contributions:
\ber \label{str_ampl} 
T_{h^\prime \lambda^\prime, h \lambda} & = & T_{h^\prime \lambda^\prime, h \lambda}^{\mathrm{non-flip}}  +  T_{h^\prime \lambda^\prime, h \lambda}^{\mathrm{flip}},
\eer
\begin{widetext}
\ber
T_{h^\prime \lambda^\prime, h \lambda}^{\mathrm{non-flip}} & = & \frac{e^2}{Q^2} \bar{u}(k^\prime,h^\prime) \gamma_\mu u(k,h)\cdot  \bar{N}(p^\prime,\lambda^\prime) \left( \cG_M \gamma^\mu -   \cF_2 \frac{P^{\mu}}{M} +  \cF_3 \frac{\gamma . K P^{\mu}}{M^2} \right) N(p,\lambda),  \label{str_ampl1}  \\
 T_{h^\prime \lambda^\prime, h \lambda}^{\mathrm{flip}} & = &\frac{e^2}{Q^2} \frac{m}{M} \bar{u}(k^\prime,h^\prime) u(k,h) \cdot \bar{N}(p^\prime,\lambda^\prime)\left( \cF_4  +  \frac{\gamma . K}{M} \cF_5  \right) N(p,\lambda)  + \frac{e^2}{Q^2} \frac{m}{M} \bar{u}(k^\prime,h^\prime) \gamma_5 u(k,h) \cdot \bar{N}(p^\prime,\lambda^\prime) \cF_6 \gamma_5 N(p,\lambda), \label{str_ampl2} \nonumber \\
\eer
\end{widetext}
where the $T$ matrix is defined as $ S = 1 + i ~T $ and $\gamma . a \equiv \gamma^\mu a_\mu$. The helicity amplitudes can be expressed in terms of the generalized form factors (FFs). Exploiting the Jacob and Wick \cite{Jacob:1959at} phase convention for spinors, the helicity amplitudes $ T_{h^\prime \lambda^\prime, h \lambda} $ in the c.m. reference frame are expressed in terms of the generalized FFs as \cite{Tomalak:2014dja}

\begin{widetext}
\ber \label{hamp}
 \Sigma_s \xi^2 \frac{T_1}{e^2} & = & 2 ( \frac{\Sigma_s Q^2}{ \Sigma_s - s Q^2} + s - M^2 - m^2 ) \cG_M - 2 (s-M^2-m^2) \cF_2 +  \frac{(s-M^2-m^2)^2}{M^2 } \cF_3 + 4 m^2 \cF_4  \nonumber \\
& & + 2 m^2 \frac{s-M^2-m^2}{M^2} \cF_5, \nonumber \\
 M \Sigma_s \xi \frac{T_2}{e^2}  & = &   2 M^2 (s - M^2 + m^2)  \cG_M - ( (s - m^2)^2 - M^4) \cF_2 + ( (s - M^2)^2 - m^4 ) \cF_3 +  2 (s + M^2-m^2) m^2 \cF_4  \nonumber \\
 & &  + 2  (s - M^2 + m^2) m^2 \cF_5,  \nonumber \\
  \Sigma_s \xi^2 \frac{T_3}{e^2}  & = & 2 (s - M^2 - m^2)   ( \cG_M - \cF_2 ) +  \frac{ (s - M^2 - m^2)^2 }{ M^2} \cF_3 + 4 m^2  \cF_4    +  2 \frac{m^2 (s - M^2 - m^2) }{ M^2 } \cF_5, \nonumber \\
  \frac{\Sigma_s }{m} \xi \frac{T_4}{e^2}  & = &  - 2 (s + M^2 - m^2)  (  \cG_M - \cF_2 )  - \frac{  ( (s - m^2)^2 - M^4)  }{ M^2} \cF_3 - 2 (s - M^2 + m^2)  \cF_4   - \frac{  ( (s - M^2)^2 - m^4 ) }{ M^2} \cF_5,  \nonumber \\
 \frac{M \Sigma_s}{m}  \frac{T_5}{e^2}  & = &   - 4 M^2 s \cG_M +   (s + M^2 - m^2)^2 \cF_2 -   ( s^2 - (M^2 - m^2)^2 ) ( \cF_3  + \cF_4)  -  \Sigma_s \cF_6 - ( s - M^2 + m^2 )^2 \cF_5,	  \nonumber \\
  \frac{M \Sigma_s}{m} \frac{T_6}{e^2}  & = &   4 M^2 s \cG_M -  (s + M^2 - m^2)^2 \cF_2 + ( s^2 - (M^2 - m^2)^2 )  ( \cF_3   + \cF_4 ) - \Sigma_s \cF_6   + ( s - M^2 + m^2 )^2 \cF_5,
\eer
\end{widetext}
with the kinematical factor $ \xi $:
\ber
 \xi & = & \sqrt{\frac{Q^2}{\Sigma_s - s Q^2}}.
 \eer
 
We consider the azimuthal angle of the scattered lepton to be $ \phi = 0 $. Notice that following the Jacob-Wick phase convention \cite{Jacob:1959at}, the azimuthal angular dependence of the helicity amplitudes is in general given by $ T_{h^\prime \lambda^\prime, h \lambda} (\theta, \phi)  = e^{ i ( \Lambda - \Lambda^\prime ) \phi } T_{h^\prime \lambda^\prime, h \lambda} (\theta, 0) $, with $ \Lambda = h - \lambda $ and $ \Lambda^\prime = h^\prime - \lambda^\prime $.

The relations of Eqs. (\ref{hamp}) can be inverted to yield the generalized FFs in terms of the helicity amplitudes as
\begin{widetext}
\ber \label{stramp}
e^2 \cG_M & = & \frac{1}{2} ( T_1 - T_3 ), \nonumber \\
 \Sigma_s e^2 \cF_2 & = & - 2 m^2 M^2  T_1  - M \left(\left(s-M^2\right)^2-m^4\right) \xi T_2    + 2 m M^2  \left(s -M^2+m^2\right) \xi T_4 - m M 
   \left(s - m^2 - M^2\right) ( T_5 - T_6) \nonumber \\
&  - & M^2 \eta(m) T_3,   \nonumber \\
 \frac{ \Sigma_s }{M^2}e^2  \cF_3 & = & - (s-M^2-m^2) T_1 - 2
   M \left(s -M^2+m^2\right) \xi  T_2   + 2 m \left(s+M^2-m^2\right) \xi  T_4 - 2 m M( T_5 - T_6) \nonumber \\
   &+& \left( \rho_3 - M^2 - m^2 \right) T_3,  \nonumber \\
 \frac{ \Sigma_s }{ M }e^2  \cF_4 & = & - M \left(s-M^2-m^2\right) T_1 - \left(\left(s -m^2\right)^2-M^4\right) \xi  T_2    +  \frac{M \left(\left(s -M^2\right)^2-m^4\right)}{ m } \xi  T_4  - \frac{  \left(s-M^2-m^2\right)^2}{2 m} ( T_5 - T_6)  \nonumber \\
& + &  M  \left( \rho_3 - M^2 - m^2 \right)  T_3, \nonumber \\
 \frac{\Sigma_s}{M^2}e^2  \cF_5 & = & 2 M^2 T_1 + 2 M \left(s+M^2-m^2\right)\xi  T_2 +  \eta(M) T_3 - \frac{\left(s-m^2\right)^2-M^4}{ m} \xi  T_4+  \frac{M \left(s-M^2-m^2\right)}{m}( T_5 - T_6),  \nonumber \\
e^2  \cF_6 & = & - \frac{M}{2m} ( T_5 + T_6 ),
\eer
\end{widetext}
with
\ber
\eta(m) & = &  \frac{ 2 m^2 \left(\Sigma_s + s Q^2 \right)+\Sigma_s Q^2}{s Q^2-\Sigma_s },   \\
 \rho_3 & = & \frac{ s \Sigma_s -  \left(M^2 - m^2 \right)^2 Q^2}{\Sigma_s  -  s Q^2 }. 
\eer

In theoretical applications, it is convenient to define also the amplitudes $\cG_1$, $\cG_2$, $\cG_3$ and $\cG_4$ through the combinations:

\ber  \label{amplitudes_G1}
 \cG_1 & = & \cG_M + \frac{\nu}{M^2} \cF_3 + \frac{m^2}{M^2} \cF_5,\label{amplitudes_G1} \\
 \cG_2 & = & \cG_M - ( 1 + \tau_P ) \cF_2 + \frac{\nu}{M^2} \cF_3, \label{amplitudes_G2} \\
  \cG_3 & = &  \frac{\nu}{M^2} \cF_3 + \frac{m^2}{M^2} \cF_5 = \cG_1 -\cG_M , \label{amplitudes_G3}   \\
 \cG_4 & = &  \cF_4 + \frac{\nu}{M^2 (1+\tau_P)} \cF_5. \label{amplitudes_G4}
\eer
 
On the one hand, the contributions to the six invariant amplitudes beyond the exchange of one photon satisfy the following model-independent relations in the forward limit, $Q^2 \to 0$ at fixed $\nu$:
\ber
 \cG_1 \left( \nu, Q^2 = 0 \right) & = & 0, \label{g1_zero_at_low_q2}  \\
 \cG_2 \left( \nu, Q^2 = 0 \right) & = & 0, \label{g2_zero_at_low_q2} \\
 \cG_4 \left( \nu, Q^2 = 0 \right) & = & 0, \label{g4_zero_at_low_q2} \\
 \left( \cF_3 + \cF_6 \right) \left( \nu, Q^2 = 0 \right) & = & \cF_4 \left( \nu, Q^2 = 0 \right). \label{f6_f4_f3_zero_at_low_q2}
\eer
We obtain these relations in Appendix \ref{app:forward_relations} analyzing the forward limit of the expressions for the helicity amplitudes in terms of invariant amplitudes, see Eqs. (\ref{hamp}). Consequently, only two among the six non-forward TPE amplitudes are independent in the forward limit.

The leading model-independent terms in the momentum transfer expansion ($Q^2 \ll m^2$) of the two-photon exchange amplitudes $\cG^{2\gamma}_1,~\cG^{2\gamma}_2,~\cG^{2\gamma}_4$ correspond to the scattering of two point charges and can be expressed as
\ber
\Re \cG^{2\gamma}_1 &\to& \frac{\alpha \pi \omega Q}{2{\bf{k}}^2} \left( 1 + \frac{m}{2M}\right),\\
\Re \cG^{2\gamma}_2 &\to& \frac{\alpha \pi \omega Q}{4 {\bf{k}}^2} \left( 1 + \frac{2m}{M}\right),\\
\Re \cG^{2\gamma}_4 &\to& - \frac{\alpha \pi M Q}{4 {\bf{k}}^2} \left(1 + \frac{m}{M} +\frac{\omega^2}{M m} \right).
\eer

On the other hand, unitarity provides constraints on the high-energy behavior, $\nu \to \infty$ at a fixed value of $Q^2$ (Regge limit), of the invariant amplitudes:
\ber
\hspace{-5.4cm} \cG_M,~\nu \cF_2,~\nu \cF_3,~\cF_4,~\cF_5,~\cF_6,  \nonumber \\
\hspace{-0.4cm} ~\cG_1,~\cG_2,~\cG_3,~\cG_4/\nu \left( \nu \to \infty, Q^2 \right)& \lesssim&  \ln^2 \nu,  \label{g1_g2_HE}
\eer
which are obtained in Appendix \ref{app:high_energy_relations}.

Performing the crossing $ \nu \to - \nu $ in the lepton (proton) line and rewriting the lepton (proton) spinors in terms of the anti-lepton (anti-proton) spinors \cite{Tomalak:2017owk}, we obtain the symmetry properties for the contributions of graphs with $ n $ exchanged photons to invariant amplitudes $\cG^{n \gamma}$:
\ber \label{crossing_relations}
 \cG^{n \gamma}_{1,2,3,M}(\nu,Q^2) & = &  (-1)^{n +1} \cG^{n \gamma}_{1,2,3,M}(-\nu,Q^2),  \\ 
 \cF^{n \gamma}_{2,5}(\nu,Q^2) & = & (-1)^{n + 1} \cF^{n \gamma}_{2,5}(-\nu,Q^2),  \\
 \cF^{n \gamma}_{3,4,6}(\nu,Q^2) & = &  (-1)^{n}  \cF^{n \gamma}_{3,4,6}(-\nu,Q^2), \\
 \cG^{n \gamma}_{4}(\nu,Q^2) & = &  (-1)^{n} \cG^{n \gamma}_{4}(-\nu,Q^2). \label{crossing_relations_last}
\eer
 
\subsection{One-photon exchange approximation}
\label{sec:OPE_approximation}

\begin{figure}[H]
\begin{center}%\centering
\includegraphics[width=.35\textwidth]{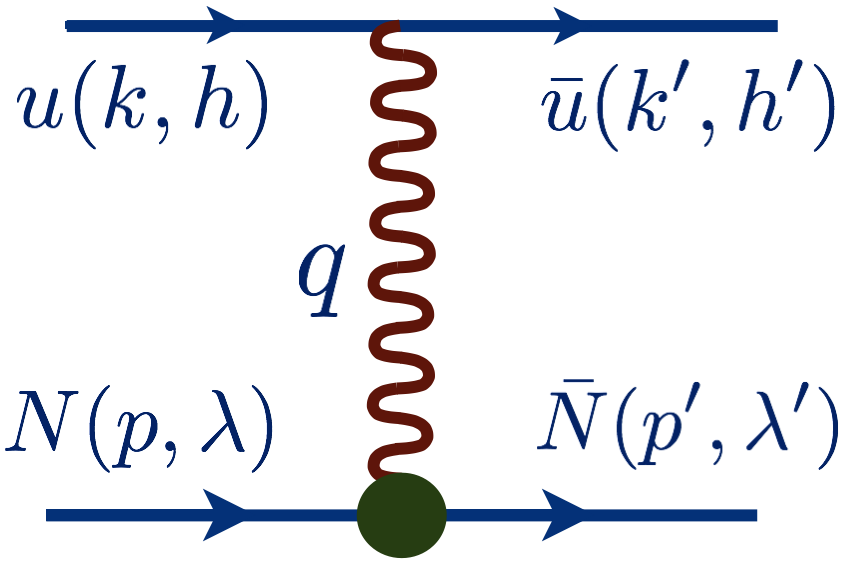}
\end{center}
\caption{Elastic lepton-proton scattering in the OPE approximation.}
\label{OPE_graph}
\end{figure}

In the one-photon exchange (OPE) approximation, the two non-zero invariant amplitudes in $ l^{-} p $ elastic scattering $ \cG_M $ and $ \cF_2 $ can be expressed in terms of the Dirac $ F_D $ and Pauli $ F_P $ FFs with the following expression for the helicity amplitude $T_{h^\prime \lambda^\prime, h \lambda}^{1\gamma}$ \cite{Foldy:1952}:
\ber \label{OPE_amplitude} 
T_{h^\prime \lambda^\prime, h \lambda}^{1\gamma}  & = &  \frac{e^2}{Q^2} \bar{u}(k^\prime,h^\prime) \gamma_\mu u(k,h) \nonumber \\
&&\hspace{-1.cm}  \times \bar{N}(p^\prime,\lambda^\prime) \left( \gamma^\mu F_D(Q^2) + \frac{i \sigma^{\mu \nu} q_\nu}{2 M} F_P(Q^2)  \right) N(p,\lambda),\nonumber \\
\eer
that is just a product of lepton and proton currents. See Fig. \ref{OPE_graph} for notations.
It is customary in experimental analysis to work with Sachs magnetic $G_M$ and electric $G_E$ FFs:
\ber  \label{Sachs_ffs}
 G_M  =  F_D + F_P , ~~~~~~~
 G_E  = F_D - \tau_P F_P,
\eer
where $ \tau_P $ is defined after Eq. (\ref{epsilon_def}). For non-relativistic systems, such as atomic nuclei, the Sachs electromagnetic proton FFs have the physical interpretation as Fourier transforms of the density of the electric charge and magnetization \cite{Soper:1972xc}. For relativistic systems, an analogous interpretation is valid only in the infinite-momentum frame \cite{Soper:1972xc}. In the OPE approximation, the invariant amplitudes defined in Eqs. (\ref{str_ampl1}) and (\ref{str_ampl2}) can be expressed in terms of the proton FFs as $ \cG^{1\gamma}_M = G_M (Q^2), ~\cF^{1\gamma}_2 = F_P (Q^2), ~\cF^{1\gamma}_3 = \cF^{1\gamma}_4 = \cF^{1\gamma}_5 = \cF^{1\gamma}_6 = 0  $. The exchange of more than one photon gives corrections of order $ O(\alpha) $, with $ \alpha = e^2/(4 \pi) \simeq 1/137 $, to all these amplitudes.

Averaging over the spin states of incoming particles and performing the sum over polarizations of outgoing particles, the unpolarized differential cross section in the OPE approximation in the laboratory frame is given by
\ber \label{OPE_xsection0}
\left( \frac{\mathrm{d} \sigma_{1 \gamma}}{\mathrm{d} \Omega} \right)_{\mathrm{lab}} = \frac{1}{256 \pi^2 M} \frac{|{\bf{k}}^\prime|}{|\bf{k}|} \frac{\sum \limits_{\mathrm{spin}} | T^{1 \gamma}|^2}{M + \omega - \omega^\prime \frac{|\bf{k}|}{|{\bf{k}}^\prime|} \cos \theta_{\mathrm{lab}} },
\eer
with the lepton solid angle $ \Omega $. We obtain in the laboratory frame:
\ber
\left( \frac{\mathrm{d} \sigma_{1 \gamma}}{\mathrm{d} \Omega} \right)_{\mathrm{lab}} \hspace{-0.2cm}= \frac{\alpha^2}{2 M}  \frac{|{\bf{k}}^\prime|}{|\bf{k}|} \frac{1}{1- \varepsilon_\mathrm{T}} \frac{G_M^2 + \frac{\varepsilon}{\tau_P} G_E^2 }{M + \omega - \omega^\prime \frac{|\bf{k}|}{|{\bf{k}}^\prime|} \cos \theta_{\mathrm{lab}} } ,
\eer
an analogue of the Rosenbluth expression \cite{Chen:2013udl,Tomalak:2014dja,Gramolin:2014pva} in agreement with Ref. \cite{Preedom:1987mx}. The unpolarized differential cross section can be equivalently written in the compact form:
\ber \label{OPE_xsection}
\frac{\mathrm{d} \sigma_{1 \gamma}}{\mathrm{d} Q^2} = \frac{\pi \alpha^2}{2 M^2 {\bf{k}}^2} \frac{ G_M^2 + \frac{\varepsilon}{\tau_P} G_E^2}{1- \varepsilon_\mathrm{T}}.
\eer

\subsection{Two-photon exchange contribution}
\label{sec:TPE_approximation}

The TPE correction to the unpolarized elastic lepton-proton scattering cross section is given by the interference between the OPE amplitude and the sum of box and crossed-box graphs with two exchanged photons. The TPE contribution $ \delta_{2 \gamma} $ at leading order in $ \alpha $ can be defined through the difference between the cross section with account of the exchange of two photons and the cross section in the $1 \gamma $-exchange approximation $ \sigma_{1 \gamma} $ as 
\ber \label{TPE_definition}
 \sigma = \sigma_{1 \gamma} \left( 1 + \delta_{2 \gamma} \right).
\eer
The leading TPE correction to the elastic $ l^{-} p $ scattering can be expressed in terms of TPE contributions to invariant amplitudes as
\ber \label{delta_TPE_massive}
\delta_{2\gamma}  & = & \frac{2}{ G_M^2 + \frac{\varepsilon}{\tau_P} G_E^2} \left\{ G_M \Re \cG^{2\gamma}_1 + \frac{\varepsilon}{\tau_P} G_E \Re \cG^{2\gamma}_2 \right. \nonumber \\ 
&+ & \left. \hspace{-0.2cm} \left( 1 - \varepsilon_\mathrm{T} \right) \left(  \frac{\varepsilon_0}{\tau_P}  \frac{\nu}{M^2}  G_E \Re \cG^{2\gamma}_4  - G_M \Re \cG^{2\gamma}_3 \right) \right\}.
\eer

Note that in the forward limit ($Q^2 \to 0$) at fixed $\nu$ \cite{Tomalak:2015hva}:
\ber \label{delta_TPE_massive_low_Q2}
\delta_{2\gamma} \to 2 \left( \Re \cG_2 + \frac{m^2}{\nu} \Re \cG_4 \right).
\eer

In this work, we follow the Maximon and Tjon prescription \cite{Maximon:2000hm} for the infrared-divergent part of the TPE contribution.  We subtract the infrared-divergent term $\delta^{\mathrm{IR}}_{2 \gamma}$ \cite{Tomalak:2014dja} corresponding with the box diagram with intermediate proton:
\ber \label{IR_delta_TPE}
\delta^{\mathrm{IR}}_{2 \gamma} & =&  \frac{2 \alpha}{\pi}  \ln \left(\frac{Q^2}{\mu^2}\right) \nonumber \\
 &\times& \hspace{-0.1cm} \left\{  \frac{s - M^2 - m^2}{\sqrt{\Sigma_s}}   \ln \frac{\sqrt{\Sigma_s}-s+(M+m)^2}{\sqrt{\Sigma_s}+s-(M+m)^2} \right. \nonumber \\  
&-& \left. \hspace{-0.15cm}   \frac{u - M^2 - m^2}{\sqrt{\Sigma_u}}  \ln \frac{(M+m)^2+\sqrt{\Sigma_u}-u}{ (M+m)^2 - \sqrt{\Sigma_u} -u} \right \},
\eer
where $ \Sigma_u = (u-(M +m)^2)(u-(M -m)^2) $ and a small photon mass $ \mu $, which regulates the infrared divergence.

According to Eqs. (\ref{g1_zero_at_low_q2}-\ref{g4_zero_at_low_q2}), the TPE correction to the unpolarized cross section vanishes in the forward limit. In the high-energy limit at fixed value of $Q^2$, corresponding with 
\ber 
\nu \to \infty,  \qquad 1 - \varepsilon_\mathrm{T} \to \left(1 + \tau_P \right) \frac{Q^2 M^2}{2\nu^2} + \mathrm{O}  \left( \frac{1}{\nu^4} \right),
\eer
the invariant amplitudes behavior is constrained by the unitarity according to Eqs. (\ref{g1_g2_HE}). Consequently, the TPE correction of Eq. (\ref{delta_TPE_massive}) vanishes in the high-energy limit if the amplitudes $ \Re \cG^{2\gamma}_1 $, $ \Re \cG^{2\gamma}_2 $ and  $ \Re \cG^{2\gamma}_4 / \nu$ vanish, which is valid for the dispersive calculation \cite{Tomalak:2014sva}, \cite{Tomalak:2017shs} and model calculations of the proton  \cite{Tomalak:2014sva}, \cite{Tomalak:2014dja} and inelastic intermediate states \cite{Tomalak:2015aoa} reflecting the odd nature of these amplitudes.

The measurement of the vanishing in OPE approximation single-spin asymmetry allows to cross-check theoretical TPE calculations. The asymmetry in the scattering of the unpolarized electrons on protons polarized normal to the scattering plane (with the proton spin $ S = \pm S_n $) is called the target normal single spin asymmetry $ A_n $ \cite{DeRujula:1972te,Pasquini:2004pv}:
\ber \label{TSSA_def}
A_n = \frac{\mathrm{d} \sigma \left( S = S_n \right) - \mathrm{d} \sigma \left( S = -S_n \right)}{\mathrm{d} \sigma \left(  S = S_n \right) + \mathrm{d} \sigma \left( S = -S_n \right)},
\eer
and the asymmetry in the interaction of electrons polarized normal to the scattering plane (with the spin direction of the initial electron: $ s = \pm s_n $) on the unpolarized target is called the beam normal single spin asymmetry $ B_n$ \cite{DeRujula:1972te,Gorchtein:2004ac}:
\ber \label{BSSA_def}
B_n = \frac{\mathrm{d} \sigma \left( s = s_n \right) - \mathrm{d} \sigma \left( s = -s_n \right)}{\mathrm{d} \sigma \left(  s = s_n \right) + \mathrm{d} \sigma \left( s = -s_n \right)}.
\eer

The asymmetries of Eqs. (\ref{TSSA_def}) and (\ref{BSSA_def}) are expressed in terms of the imaginary parts of TPE amplitudes at leading order in $ \alpha $ as
\ber
A_n & = & \sqrt{\frac{2\varepsilon_\mathrm{T} \left( 1 + \varepsilon_\mathrm{T} \right)}{\tau_P}} F \left\{ \frac{ G_E \Im \cG^{2 \gamma}_1 - G_M \Im \cG^{2 \gamma}_2   }{ G^2_M  + \frac{\varepsilon}{\tau_P} G^2_E} \right. \nonumber \\
&& \left. \hspace{-0.65cm} -  \frac{1 + \tau_P}{\nu} \left( \frac{\tau_P M^2 G_E   \Im \cF^{2 \gamma}_3 + m^2 G_M \Im \cG^{2 \gamma}_4 }{ G^2_M  + \frac{\varepsilon}{\tau_P} G^2_E} \right) \right\} ,   \label{TSSA_massive}\\
B_n & = & - \frac{m}{M} \frac{\sqrt{1+\tau_P}}{\tau_P} \sqrt{2   \varepsilon_\mathrm{T} \left(1-\varepsilon_\mathrm{T} \right)} \nonumber \\
&&  \hspace{-0.85cm}\times\frac{ \left( G_E + \tau_P G_M \right)\Im  \cG^{2\gamma}_4 + \tau_P G_M  \Im  \left(\cF^{2 \gamma}_3 - \cF^{2 \gamma}_4  \right) }{ G^2_M  + \frac{\varepsilon}{\tau_P} G^2_E },  \label{BNSSA_massive}
\eer
where we introduced a kinematical factor $F$:
\ber \label{factor_f}
F = \sqrt{1 + 2\varepsilon_0 \left( \frac{1 -\varepsilon_\mathrm{T}}{ 1 + \varepsilon_\mathrm{T} } \right) },
\eer
which is equal to 1 in the lepton massless limit.

Note that the amplitude $ \cG^{2\gamma}_4 $ introduced in Eq. (\ref{amplitudes_G4}) appears also in the expression for the unpolarized cross section of Eq. (\ref{delta_TPE_massive}). The contribution to $ A_n,~B_n $ and $ \delta_{2 \gamma } $ which is linear in the amplitude $ \cF^{2\gamma}_6 $ vanishes \cite{Tomalak:2014dja,Gakh:2014zva}. The amplitude $ \cF^{2\gamma}_6 $ only shows up in double polarization observables, which are influenced by real parts of TPE amplitudes. 

In the following, we consider the polarization transfer observables from the longitudinally polarized electron to the recoil proton accounting for the leading TPE contributions. The longitudinal polarization transfer asymmetry is defined as
\ber
P_l = \frac{\mathrm{d} \sigma \left( h = +,~\lambda'=+ \right) - \mathrm{d} \sigma \left( h = +,~\lambda'=- \right)}{\mathrm{d} \sigma \left( h = +,~\lambda'=+ \right) + \mathrm{d} \sigma \left( h = +,~\lambda'=- \right)},
\eer
and the transverse polarization transfer asymmetry is given by
\ber
P_t = \frac{\mathrm{d} \sigma \left( h = +,~S'=S_\perp \right) - \mathrm{d} \sigma \left( h = +,~S'=-S_\perp \right)}{\mathrm{d} \sigma \left( h = +,~S'=S_\perp \right) + \mathrm{d} \sigma \left( h = +,~S'= - S_\perp \right)}, \nonumber \\
\eer
with the spin direction of the recoil proton $ S' = \pm S_\perp $ in the scattering plane transverse to its momentum direction.

The transverse polarization transfer observable $ P_t $ relative to the Born result $ P^{\mathrm{Born}}_t $ is given by
\ber
 \frac{P_t}{P^{\mathrm{Born}}_t} & = &  1 + \delta_t = 1   - \delta_{2 \gamma} + \frac{\Re \cG_M^{2 \gamma}}{G_M} + \frac{\Re \cG_2^{2 \gamma}}{G_E}  \nonumber \\
 &+&  \frac{m^2}{ M \omega} \frac{ \left( 1 + \tau_P \right)  \Re \cG_4^{2 \gamma} + \tau_P \Re \cF_6^{2 \gamma}}{G_E}, \label{polarization_observables2xx}
\eer
with a relative correction $\delta_t$ and the leading-order expression:
\ber
P^{\mathrm{Born}}_t &=& -  \sqrt{\frac{2 \varepsilon_\mathrm{T} \left( 1- \varepsilon_\mathrm{T}  \right)}{\tau_P}} \frac{\omega}{|\vec{k}|} \frac{ G_E G_M}{G_M^2 + \frac{\varepsilon}{\tau_P} G_E^2} .
\eer

The longitudinal polarization transfer observable $ P_l $ relative to the Born result $ P^{\mathrm{Born}}_l $ is given by
\ber
 \frac{P_l}{P^{\mathrm{Born}}_l} & = & 1 + \delta_l =  1   - \delta_{2 \gamma}  + 2 \frac{\Re \cG_M^{2 \gamma}}{G_M} \nonumber \\
 &+& \frac{ 2 \varepsilon_\mathrm{T}}{1 + \varepsilon_\mathrm{T}} \frac{1}{1 + a F}   \frac{ \Re \cG_3^{2 \gamma} + \tau_P \Re \cF_3^{2 \gamma}  }{G_M} \nonumber \\
 &-& \frac{1-\varepsilon_\mathrm{T}}{1+\varepsilon_\mathrm{T}}   \frac{ 1 +\frac{ \omega}{M} }{ 1 + a F}  \frac{m^2}{M^2} \frac{ G_E \Re \cF_6^{2 \gamma}}{ \left( 1 + \tau_P\right) \tau_P G_M^2}, \label{polarization_observables2xx}
\eer
with a relative correction $\delta_l$, and where the kinematical parameter $a$ is defined as
\ber
a= \sqrt{\frac{\tau_P}{1+\tau_P}} \sqrt{\frac{1-\varepsilon_\mathrm{T}}{1+\varepsilon_\mathrm{T}}}.
\eer
The leading-order expression is given by
\ber
P^{\mathrm{Born}}_l &=&  \sqrt{1 - \varepsilon_\mathrm{T}^2} \frac{1 + a F}{a + F} \frac{\omega}{|\vec{k}|} \frac{ G^2_M}{G_M^2 + \frac{\varepsilon}{\tau_P} G_E^2}.
\eer

The ratio of polarization transfer observables $P_t/P_l$ can be expressed as
\ber
\frac{P_t}{P_l} & = & - \sqrt{\frac{2 \varepsilon_\mathrm{T}}{\tau_P \left(1+\varepsilon_\mathrm{T} \right)}} \frac{a + F}{1 + a F} \frac{G_E}{G_M} \left( 1 + \delta_t - \delta_l \right).
\eer

The other double polarization transfer observables $A_t$ and $A_l$ with a polarized target (in the same direction as a recoil proton in $P_t $ and $P_l$) are related to the polarization transfer observables by  
\ber
A^{\mathrm{Born}}_t & = & P^{\mathrm{Born}}_t, \\
A^{\mathrm{Born}}_l & = & - P^{\mathrm{Born}}_l.
\eer
The relative TPE corrections $\delta_l $ and $\delta_t$ from amplitudes $\cG^{2\gamma}_M,~\cF^{2\gamma}_2,~\cF^{2\gamma}_3,~\cF^{2\gamma}_4,~\cF^{2\gamma}_5$ are the same for the target polarization asymmetries and polarization transfer observables. However, the contribution from $\cF_6^{2\gamma}$ has an opposite sign.

\section{Dispersion relation formalism in muon-proton scattering}
\label{sec3}

In this Section, we describe the dispersion relation formalism to evaluate the two-photon exchange correction to all six invariant amplitudes in the elastic muon-proton scattering, i.e., when including the lepton mass terms. Unitarity relations allow us to unambiguously reconstruct imaginary parts of TPE amplitudes for the contribution of the individual channel. The resulting correction is given by a sum of all intermediate states. In this work, we discuss the leading elastic contribution. We reconstruct the real parts of the amplitudes which enter the cross-section correction using fixed-$Q^2$ dispersion relations. For the amplitude $\cF_4$, a once-subtracted dispersion relation is required, whereas the real parts of $\cG_1,~\cG_2,~\cF_3$ and $\cF_5$ can be reconstructed using unsubtracted DRs. Finally, we describe the way to predict the TPE correction $\delta_{2\gamma}$ at different values of $\nu$ relying on the known correction at some point $\nu_0$.

\subsection{Unitarity relations}
\label{sec:unitarity}

We obtain the imaginary parts of invariant amplitudes exploiting the unitarity equation for the scattering matrix $S$:
\ber
 S^+ S  =  1,\qquad T^+ T = i ( T^+ - T).
\eer
In the c.m. reference frame, we reconstruct the imaginary part of the TPE helicity amplitude $ \Im T^{2 \gamma}_{h^\prime \lambda^\prime,h \lambda} $ by the phase-space integration of the product of OPE amplitudes from the initial to intermediate state $ T^{1 \gamma}_{ \mathrm{hel} , h \lambda}  $ and from the intermediate state to final state $ T^{1 \gamma}_{h^\prime \lambda^\prime, \mathrm{hel}} $:

\ber \label{unitarity_relations}
\Im T^{2 \gamma}_{h^\prime \lambda^\prime,h \lambda} & = & \frac{1}{2} \sum \limits_{n,\mathrm{hel} } \prod \limits_{i=1}^n \int \frac{\mathrm{d}^3 \bf{q}_i}{(2 \pi)^3} \frac{1}{2 E_i} ( T^{1 \gamma}_{\mathrm{hel}  , h^\prime \lambda^\prime} )^{*} T^{1 \gamma}_{ \mathrm{hel} , h \lambda} \nonumber \\
&& (2 \pi)^4 \delta^4 (k+p-\sum_i q_i),
\eer
where the sum goes over all possible number $ n $ of intermediate particles with momenta $ q_i = ( E_i, \bf{q}_i ) $ and all possible helicity states (denoted as "$ \mathrm{hel} $"). Unitarity relations allow us to relate the imaginary part of the TPE amplitude to the experimental OPE input in a model-independent way.

In the following, we describe the kinematics of the intermediate state in the lepton-proton c.m. reference frame, as we exploit this frame relating the lepton-proton helicity amplitudes to invariant amplitudes in Section \ref{sec:HA_and_generalized_FFs}.

\begin{figure}[H]
\begin{center}%\centering
\includegraphics[width=0.45\textwidth]{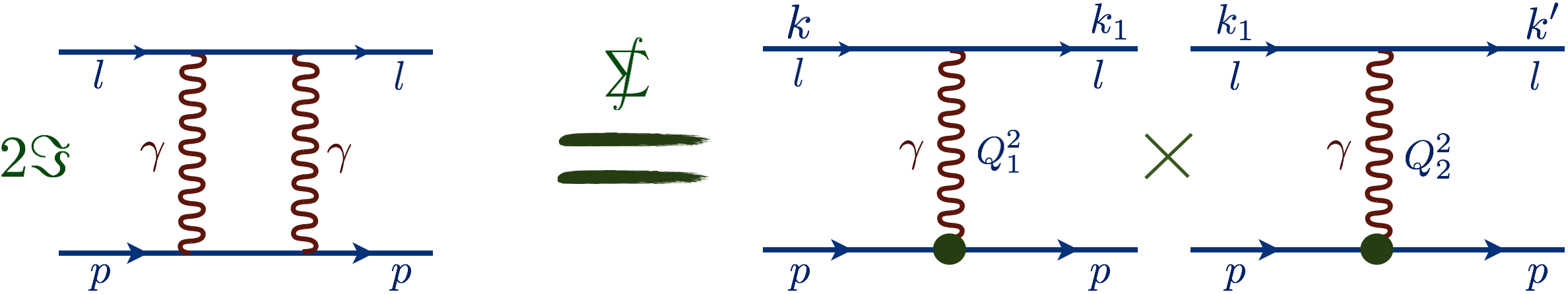}
\end{center}
\caption{Unitarity relations for the case of the elastic intermediate state contribution.}
\label{unitarity}
\end{figure}

The unitarity relations are represented in Fig.~\ref{unitarity} for the elastic intermediate state.

For an arbitrary intermediate state with the squared invariant mass $W^2 = (p+k-k_1)^2$, the intermediate lepton energy $ \omega_1$ and momentum $ | {\bf{k}}_1 | $ are given by
\ber \label{lepton_momentum}
 \omega_1 =  \frac{s - W^2 + m^2}{2\sqrt{s}} , \qquad  | {\bf{k}}_1 | = \frac{ \sqrt{\Sigma(s, W^2, m^2)} }{2 \sqrt{s}}.
\eer
For a proton intermediate state, the intermediate lepton momentum is obtained by the substitution $W \to M$ resulting in $\omega_1 = \omega_{\mathrm{cm}}$ and $| {\bf{k}}_1 | = | {\bf{k}}_{\mathrm{cm}} |$.

The lepton initial ($k$), intermediate ($k_1$) and final ($k^\prime$) momenta are given by
\ber \label{kinematics_unitarity}
 k & = & (\omega_{\mathrm{cm}}  ,0, 0, | {\bf{k}}_{\mathrm{cm}} |),  \label{lepton_initial} \\
 k_1 & = & (\omega_1  , | {\bf{k}}_1 | \sin \theta_1 \cos \phi_1 , | {\bf{k}}_1 | \sin \theta_1 \sin \phi_1 , \nonumber \\
 &&  | {\bf{k}}_1 | \cos \theta_1), \label{lepton_intermediate} \\ 
 k^\prime & = & (\omega_{\mathrm{cm}}  , | {\bf{k}}_{\mathrm{cm}} | \sin \cma , 0, | {\bf{k}}_{\mathrm{cm}} | \cos \cma),  \label{lepton_final}
\eer 
with the intermediate lepton angles $ \theta_1 $ and $ \phi_1 $.

We also introduce the relative angle $\theta_2$ between the 3-momenta of intermediate and final leptons as 
\ber
 {\bf{k}}_1 \cdot {\bf{k}}^\prime =  | {\bf{k}}_{\mathrm{cm}} | | {\bf{k}}_1 | \cos \theta_2,
 \eer
 with $ \cos \theta_2 = \cos \cma \cos \theta_1 + \sin \cma \sin \theta_1 \cos \phi_1 $.

The squared virtualities of the exchanged photons $ Q^2_1=-(k-k_1)^2$ and $Q^2_2=-(k^\prime-k_1)^2$ can be expressed as
\ber \label{Q21_Q22_ellipse}
Q^2_{1,2} & = &  \frac{\left( s - M^2 + m^2 \right) \left(s - W^2 + m^2 \right) - 4 m^2 s }{2 s}  \nonumber \\
&-& \frac{\sqrt{ \Sigma \left( s, M^2, m^2 \right) \Sigma \left( s, W^2, m^2 \right)}}{2s} \cos \theta_{1,2}.
\eer

Now, we discuss the unitarity relations of Eq. (\ref{unitarity_relations}). We follow Refs. \cite{Pasquini:2004pv,Tomalak:2016vbf,Tomalak_PhD} generalizing all expressions to the case of massive leptons.

For the hadronic intermediate state, we include the hadronic phase-space integration and the sum over hadron polarizations in Eq. (\ref{unitarity_relations}) into the hadronic tensor $ W^{\mu \nu}$ and express the imaginary part of the TPE helicity amplitude as
\ber \label{piN_unitarity}
\Im T^{2 \gamma}_{h^\prime \lambda^\prime, h \lambda} \hspace{-0.15cm} & = &\hspace{-0.12cm} \frac{e^4}{2} \hspace{-0.07cm} \int \hspace{-0.14cm} \frac{\mathrm{d}^3 \vec{k}_1}{ (2 \pi)^3 2 \omega_1} \hspace{-0.05cm}  \frac{\bar{u}(k^\prime,h^\prime) \gamma_\mu \left( \gamma.k_1 + m \right) \gamma_\nu u(k,h)}{Q^2_1 Q^2_2}  \nonumber \\ 
 &\times& \hspace{-0.12cm} \bar{N}(p^\prime,\lambda^\prime) W^{\mu \nu} \left( p, p^\prime, k_1   \right) N(p,\lambda).
\eer
The imaginary parts of the invariant amplitudes are given by relations of Eqs. (\ref{stramp}).

The proton intermediate state contribution to the hadronic tensor is given by
\ber
W^{\mu \nu} \left( p, p^\prime, k_1   \right) & = & \gamma_0 \left(J^{\mu}_{p}(p_1, p^\prime ) \right)^\dagger \gamma_0  \left( \gamma.p_1 + M \right)  J_{p}^\nu (p_1, p) \nonumber \\
&\times& 2 \pi \delta(W^2 - M^2),
\eer
with the proton momentum $p_1 = p + k - k_1$ and electromagnetic current $J^\mu_{p} $ from Eq. (\ref{OPE_amplitude}):
\ber \label{proton_current}
J_{p}^\mu (p_1, p)  =  G_M  \gamma^\mu - F_2  \frac{p^{\mu} + p^\mu_1}{2 M}.
\eer
In the following, we exploit the dipole form for the proton form factors:
\ber \label{dipole_FFs}
 G_E (Q^2) = \frac{1}{\left( 1 + \frac{Q^2}{\Lambda^2} \right)^2}, \quad G_M (Q^2)  = \frac{\mu_P}{\left( 1 + \frac{Q^2}{\Lambda^2} \right)^2},
\eer
with the proton magnetic moment $\mu_P \approx 2.793$ and hadronic scale $ \Lambda^2 = 0.71 ~\mathrm{GeV}^2 $.

The imaginary part of the elastic contribution can be also expressed as an integral over the product of OPE helicity amplitudes:
\ber \label{unitarity_relations_proton_state}
\Im T^{2 \gamma}_{h^\prime \lambda^\prime,h \lambda}  = \frac{\sqrt{\Sigma_s}}{64 \pi^2 s} \sum \limits_{\tilde{h} \tilde{\lambda}} \int \mathrm{d} \Omega_1 \left( T^{1 \gamma}_{\tilde{h} \tilde{\lambda}, h^\prime \lambda^\prime} \right)^{*} T^{1 \gamma}_{\tilde{h} \tilde{\lambda}, h \lambda}.
\eer
We will exploit Eq. (\ref{unitarity_relations_proton_state}) as a numerical cross check in the following.

We checked that the numerical calculations of the imaginary parts of the invariant amplitudes are in agreement with theoretical predictions for the target and beam normal single spin asymmetries $ A_n $ and $B_n$ \cite{Pasquini:2004pv,DeRujula:1972te}, given by Eqs. (\ref{TSSA_massive}) and (\ref{BNSSA_massive}). The resulting amplitudes are in agreement with the low-momentum transfer
 limit of Eqs. (\ref{g1_zero_at_low_q2}-\ref{f6_f4_f3_zero_at_low_q2}). Moreover, the imaginary parts of all invariant TPE amplitudes are in exact agreement with the model calculation of the proton intermediate state contribution of Ref. \cite{Tomalak:2014dja}, see Appendix \ref{app:hm_vs_drs} for some details.

\begin{figure}[H]
\begin{center}%\centering
\includegraphics[width=0.45\textwidth]{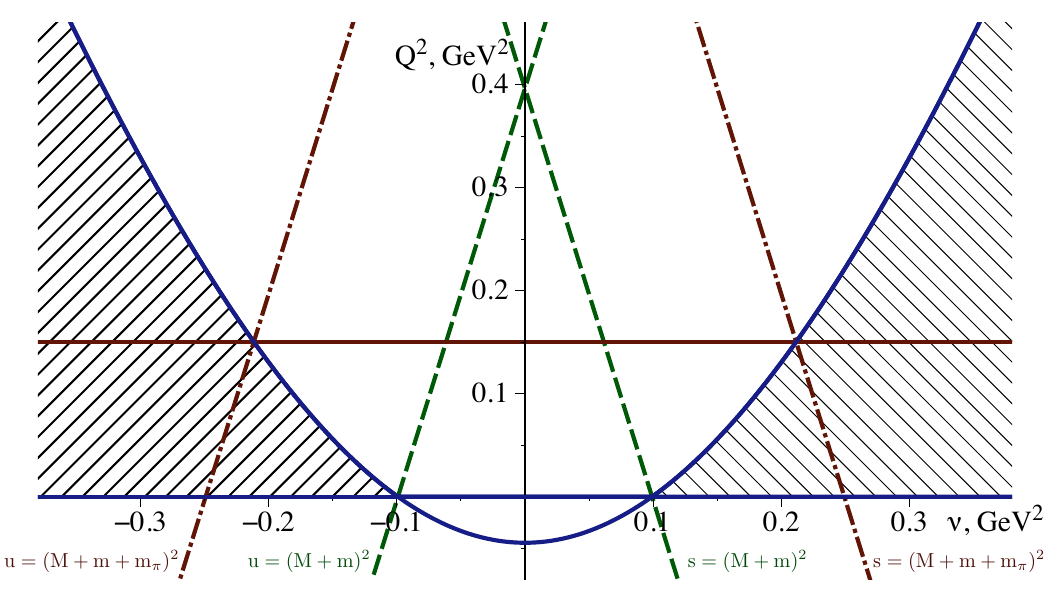}
\end{center}
\caption{Physical and unphysical regions of the kinematical variables $ \nu $ and $ Q^2 $ (Mandelstam plot) for the elastic muon-proton scattering. The hatched blue region corresponds to the physical region, the dashed green lines give the elastic threshold positions, the dashed-dotted red lines give the inelastic threshold positions. The horizontal red curve indicates the line at fixed $ Q^2 $ along which the dispersive integrals are evaluated. For $ Q^2 \gtrsim 0.4 ~\mathrm{GeV}^2 $ ($ Q^2 \gtrsim 1 ~\mathrm{GeV}^2 $) the s- and u-channel elastic (pion-nucleon) cuts overlap. }
\label{mandelstam}
\end{figure}

To evaluate the dispersive integral at a fixed value of momentum transfer $ Q^2 $, we have to know the imaginary parts of the invariant amplitudes from the production threshold in energy upwards. When evaluating the imaginary parts through the unitarity relations as a phase-space integration, it only covers the "physical" region of the dispersive integrand. However, the invariant amplitudes also have an imaginary part outside the physical domain as long as one is above the elastic threshold and thus require an analytical continuation outside the physical domain. To illustrate the physical and unphysical regions,  we show in Fig. \ref{mandelstam} the Mandelstam plot for the elastic muon-proton scattering.

The boundary of the physical region is given by the hyperbola:
\ber \label{nuphys_massive}
\nu = \nu_{\mathrm{ph}} \equiv M m \sqrt{1+\tau_P} \sqrt{1+\tau_l},
\eer
where $ \tau_l $ and $ \tau_P $ were defined after Eq. (\ref{epsilon_def}). Therefore, the evaluation of the dispersive integral for the elastic intermediate state contribution requires information from the unphysical region for any $ Q^2 > 0 $. We perform the analytical continuation for the elastic intermediate state by the countour-deformation method of Ref.~\cite{Tomalak:2014sva}, which was proven to be exact for parametrizations as a sum of dipoles or monopoles and, therefore, this method is valid in our calculation. The intersection between the backward angle branch of the hyperbola of Eq. (\ref{nuphys_massive}) and the line $ s = (M+m+m_\pi)^2$ describing the first pion-nucleon inelastic threshold corresponds with:
\ber
Q_{\mathrm{th}}^2 &=& \frac{ (2 M+ 2  m + m_\pi) (2 M + m_\pi ) (2 m + m_\pi ) m_\pi}{( M + m + m_\pi )^2} \nonumber \\
&\simeq& 0.150 ~ \mathrm{GeV}^2,
\eer
 (indicated by the red horizontal line in Fig. \ref{mandelstam}), where $m_\pi$ denotes the pion mass. Therefore, the kinematically allowed momentum transfer region of the MUSE experiment $ Q^2 < 0.116~\mathrm{GeV}^2 <  Q_{\mathrm{th}}^2 $ does not require an analytical continuation into the unphysical region for inelastic contributions.

\subsection{Dispersion relations}
\label{sec:dispersion}

Assuming the analyticity of invariant amplitudes, we obtain the real parts of TPE amplitudes by evaluating dispersive integrals.

According to the high-energy behavior of invariant amplitudes of Eqs.~(\ref{g1_g2_HE}), the odd TPE amplitude $ \cF^{2 \gamma}_2 $ and the even amplitude $ \cF^{2 \gamma}_3 $ vanish in the Regge limit $\nu\rightarrow\infty$, $Q^{2}/\nu \rightarrow 0$. Such high-energy behavior allows us to neglect the contribution from the infinite contour considering the Cauchy's theorem and to write down the unsubtracted DRs at a fixed value of the momentum transfer $ Q^2 $. For other odd in $\nu$ amplitudes: $~\cF^{2 \gamma}_5, ~\cG^{2 \gamma}_1, ~\cG^{2 \gamma}_2$ entering Eq. (\ref{delta_TPE_massive}), the possible contribution from the infinite contour vanishes due to the odd property under the reflexion $\nu \to - \nu$. Consequently, we are allowed to write down the unsubtracted dispersion relations for the odd amplitudes $ {\cal{G}}^{\mathrm{\mathrm{odd}}}$: $~\cF^{2 \gamma}_2,~\cF^{2 \gamma}_5, ~\cG^{2 \gamma}_1, ~\cG^{2 \gamma}_2$ and for the even amplitude $ \cF^{2 \gamma}_3 $:
\ber
 \label{oddDR}
 \Re {\cal{G}}^{\mathrm{\mathrm{odd}}}(\nu, Q^2) & = & \frac{2 \nu}{\pi} \fint \limits^{~ \infty}_{\nu_{\mathrm{thr}}} \frac{\Im {\cal{G}}^{\mathrm{\mathrm{odd}}} (\nu^\prime, Q^2)}{{\nu^\prime}^2-\nu^2}  \mathrm{d} \nu^\prime, \\
 \label{evenDR}
 \Re  {\cal{F}}^{2 \gamma}_3 (\nu, Q^2) & = & \frac{2}{\pi} \fint \limits^{~ \infty}_{\nu_{\mathrm{thr}}}  \nu^\prime \frac{\Im  {\cal{F}}^{2 \gamma}_3  (\nu^\prime, Q^2)}{{\nu^\prime}^2-\nu^2}  \mathrm{d} \nu^\prime,
\eer
where the imaginary part is taken from the $s$-channel discontinuity only. The elastic threshold position is given by $ \nu_{\mathrm{thr}} = M m - Q^2 / 4$, while the pion-nucleon intermediate states start to contribute from:
\ber
 \nu^{\pi N}_{\mathrm{thr}} =  M m -Q^2/4 + (M+m) m_\pi  + m_{\pi}^2/2.
\eer
The kinematical points of MUSE ($\nu = M  \omega - Q^2 / 4$) are below the pion-production threshold.

According to the high-energy relations of Eqs. (\ref{g1_g2_HE}), one cannot write down the unsubtracted DR for the even amplitude $ \cF_4^{2 \gamma}$. Note that the other even amplitude $\cF_6^{2 \gamma}$ does not contribute to the unpolarized cross section at leading order.

\subsection{Subtracted dispersion relation formalism}
\label{sec:subtracted}

One can apply Cauchy's theorem for the amplitude $\cF_4^{2\gamma}$ subtracted at a point $\nu_0$: $ \cF_4^{2 \gamma} (\nu,~Q^2) - \cF_4^{2 \gamma} (\nu_0,~Q^2)$. Deforming the integration contour to infinity, we obtain the once-subtracted dispersion relation:
\ber \label{once_subtracted}
 \Re \cF_4^{2 \gamma} (\nu, Q^2)& = &  \Re \cF_4^{2 \gamma} (\nu_0, Q^2) \nonumber \\
&& \hspace{-1.25cm} + \frac{2 \left( \nu^2 - \nu_0^2 \right)}{\pi} \fint \limits^{~ \infty}_{\nu_{\mathrm{thr}}} \frac{ \nu^\prime  \Im \cF_4^{2 \gamma}  (\nu^\prime, Q^2)  \mathrm{d} \nu^\prime}{\left( {\nu^\prime}^2-\nu^2 \right)\left({\nu^\prime}^2-\nu_0^2 \right )}. 
\eer

The subtraction in the dispersion relation analysis corresponds with the introduction of a counterterm in the effective field theory. The counterterm near the structure $\bar{u} u \bar{N} N$ and its effect on the Lamb shift in muonic hydrogen and elastic muon-proton scattering was studied in Ref. \cite{Miller:2012ne}.

We have to fix the subtraction function $ \Re \cF_4^{2 \gamma} (\nu_0, Q^2)$ in order to make a DR prediction. In this work, we exploit the model result for $ \delta_{2 \gamma} (\nu_0, Q^2)$ of Ref. \cite{Tomalak:2015hva}, which is expected to describe the TPE correction at small momentum transfer and energy of the MUSE experiment. We separate the contribution from the amplitude $\cF_4^{2 \gamma} $ to the TPE correction of Eq. (\ref{delta_TPE_massive}) as
\ber
\hspace{-0.05cm}\delta_{2\gamma} \left( \nu,~Q^2 \right) \hspace{-0.05cm}= \hspace{-0.05cm}\delta^0_{2\gamma} \left( \nu,~Q^2 \right) \hspace{-0.05cm}+ \hspace{-0.05cm}f\left( \nu,~Q^2 \right) \Re \cF_4^{2 \gamma} \left( \nu,~Q^2 \right)\hspace{-0.08cm},
\eer
with
\ber
f\left( \nu,~Q^2 \right) =  \frac{2  \left( 1 - \varepsilon_\mathrm{T} \right) G_E}{ G_M^2 + \frac{\varepsilon}{\tau_P} G_E^2}  \frac{\varepsilon_0}{\tau_P} \frac{\nu}{M^2}.
\eer
The remaining part of the cross-section correction $\delta^0_{2\gamma}$ is given by
\ber \label{delta_TPE_massive0}
\delta^0_{2\gamma} \left( \nu,~Q^2 \right)  & = & \frac{2}{ G_M^2 + \frac{\varepsilon}{\tau_P} G_E^2} \left\{ G_M \Re \cG^{2\gamma}_1 + \frac{\varepsilon}{\tau_P} G_E \Re \cG^{2\gamma}_2 \right. \nonumber \\ 
&& \left. \hspace{-0.85cm} + \left( 1 - \varepsilon_\mathrm{T} \right)\left( \frac{\varepsilon_0}{\tau_P}  \frac{\nu^2 G_E \Re \cF^{2\gamma}_5}{M^4 \left( 1+\tau_P\right)}    - G_M \Re \cG^{2\gamma}_3 \right) \right\}, \nonumber \\
\eer
and is evaluated using unsubtracted DRs. Assuming that the leading TPE contributions are accounted for in our calculations, we can extract the one unknown amplitude $ \Re \cF_4^{2 \gamma} (\nu_0, Q^2)$ from the known cross-section correction  $\delta^{\mathrm{ref}}_{2\gamma} (\nu_0, Q^2)$ as
\ber
\Re \cF_4^{2 \gamma} (\nu_0, Q^2) = \frac{\delta^{\mathrm{ref}}_{2\gamma} (\nu_0, Q^2)-\delta^0_{2\gamma} (\nu_0, Q^2)}{f\left( \nu_0,~Q^2 \right)}.
\eer
Using the subtracted DR of Eq. (\ref{once_subtracted}), we can then predict the cross-section correction for other values of $ \nu$.

\section{Results and discussion}
\label{sec5}

In this Section, we provide our predictions for the TPE correction in MUSE kinematics within the subtracted DR formalism.

Although we are not allowed to write down the unsubtracted dispersion relation for the amplitude $\cF_4$, it is instructive to compare the unsubtracted DR prediction to the model evaluations of the TPE correction since the dispersive integral for the model calculation is convergent.

\begin{figure}[H]
\begin{center}%\centering
\includegraphics[width=0.45\textwidth]{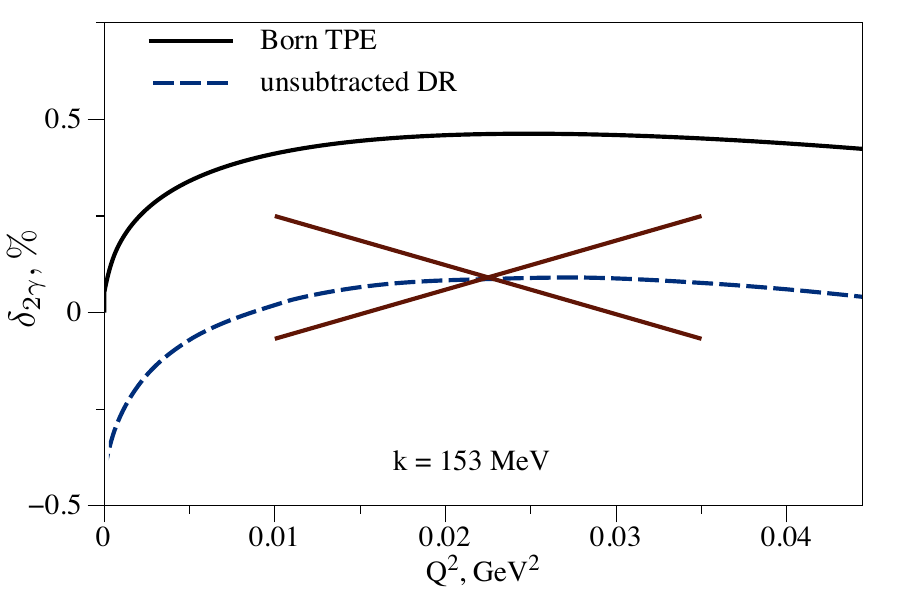}
\end{center}
\caption{TPE correction to the unpolarized elastic $ \mu^{-}p $ cross section evaluated within the unsubtracted DR framework (blue dashed curve). It is compared to the evaluation in the box diagram model (Born TPE) shown by the black solid curve.}
\label{tpe_FFs_DR0}
\end{figure}

In Fig. \ref{tpe_FFs_DR0}, we show the prediction for the elastic contribution to $ \delta_{2 \gamma} $ within unsubtracted DRs and compare it with the box graph model calculation of Ref. \cite{Tomalak:2014dja}, which is denoted as Born TPE in Fig. \ref{tpe_FFs_DR0}, for one of the MUSE beam energies. The unsubtracted DR result is significantly below the model prediction. The difference between both evaluations is mainly given by the amplitude $ \cF^{2 \gamma}_4$, see Appendix \ref{app:hm_vs_drs} for a more detailed comparison. Moreover, the unsubtracted DR evaluation of only the elastic intermediate state TPE in the forward limit yields: $ \cG^{2\gamma}_4 \left( \nu, Q^2 \to 0 \right) \neq 0$, in contradiction to the constraint of Eq. (\ref{g4_zero_at_low_q2}). Consequently, such calculation does not satisfy the expected vanishing low-$Q^2$ behavior of the cross-section correction. This violation is due to the presence of non-zero Pauli coupling in the photon-proton-proton vertex. It generates a constant term at infinity for the amplitude $\cF_4$, which has to be evaluated by the subtracted DR with a $ Q^2 $-dependent subtraction function. The latter renormalizes the effects of the momentum-dependent Pauli coupling in a proper way. MUSE will be able to provide measurements of the subtraction function for all three beam momenta in the kinematical region $ 0.0052~\mathrm{GeV}^2 < Q^2 < 0.027~\mathrm{GeV}^2$.

\begin{figure}[H]
\begin{center}%\centering
\includegraphics[width=0.45\textwidth]{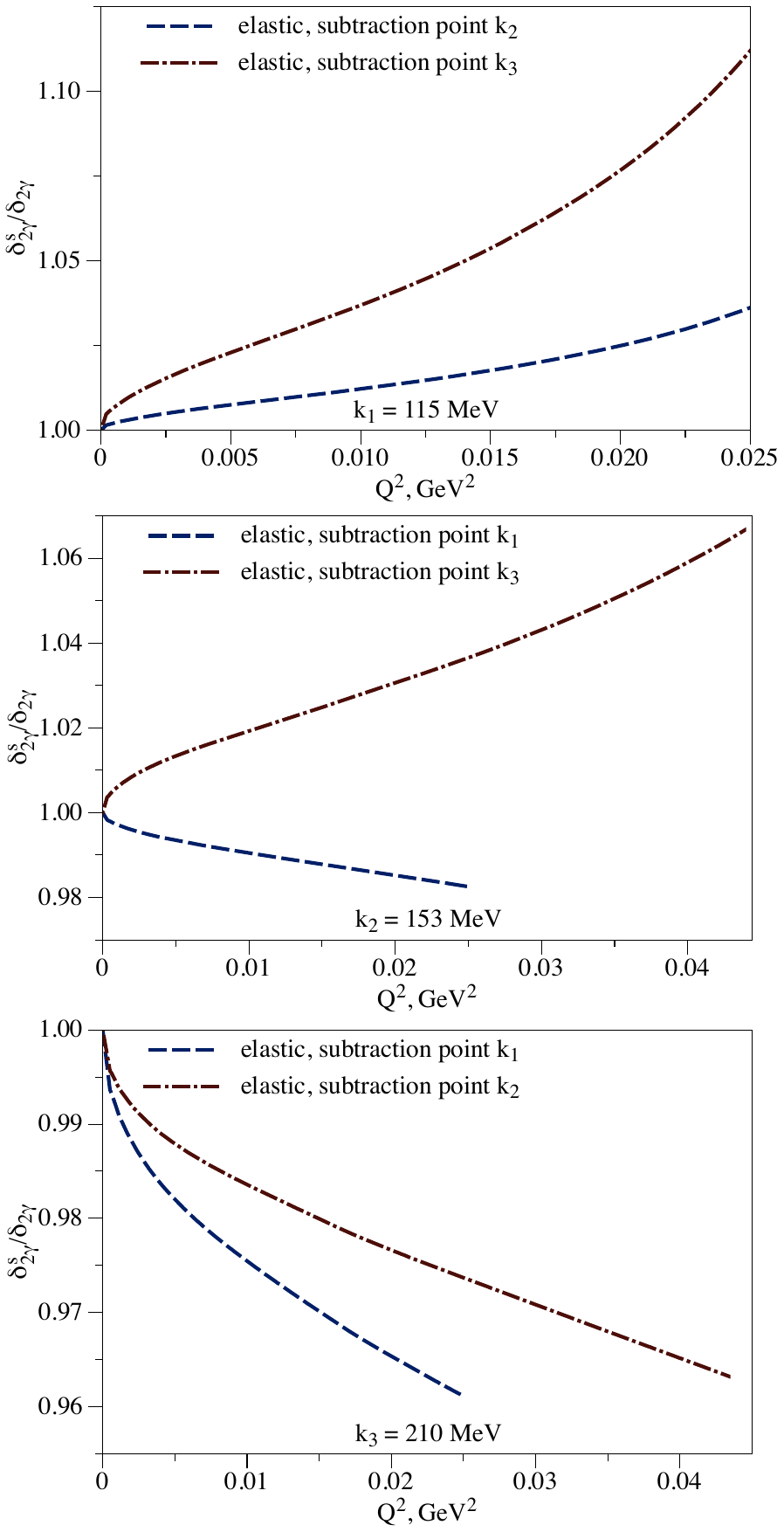}
\end{center}
\caption{The ratio of the TPE correction to the unpolarized elastic $ \mu^{-}p $ cross section within the subtracted DR framework to the correction of Ref. \cite{Tomalak:2015hva}. The results are shown by the blue dashed and red dashed-dotted curves corresponding to different subtraction points with the MUSE beam momenta: $ k_1 = 115~\mathrm{MeV},~k_2 = 153~\mathrm{MeV},~k_3 = 210~\mathrm{MeV}$.}
\label{tpe_subtracted_elastic}
\end{figure}

In the absence of the data, we take the subtraction point corresponding to MUSE beam energy from the total TPE estimate of Ref. \cite{Tomalak:2015hva}, which simulates the analysis of forthcoming data. In the subtracted dispersion relation approach, we account only for the leading elastic TPE contribution. On the plots in Fig. \ref{tpe_subtracted_elastic}, we show the ratio of our TPE prediction to the model calculation for the total TPE correction of Ref. \cite{Tomalak:2015hva}.
We notice from Fig. \ref{tpe_subtracted_elastic} that in the range of the MUSE kinematics the TPE correction for the elastic intermediate state within subtracted DRs agrees within 10$~\%$ of its value with the result of the near-forward model calculation.

\begin{figure}[H]
\begin{center}%\centering
\includegraphics[width=0.45\textwidth]{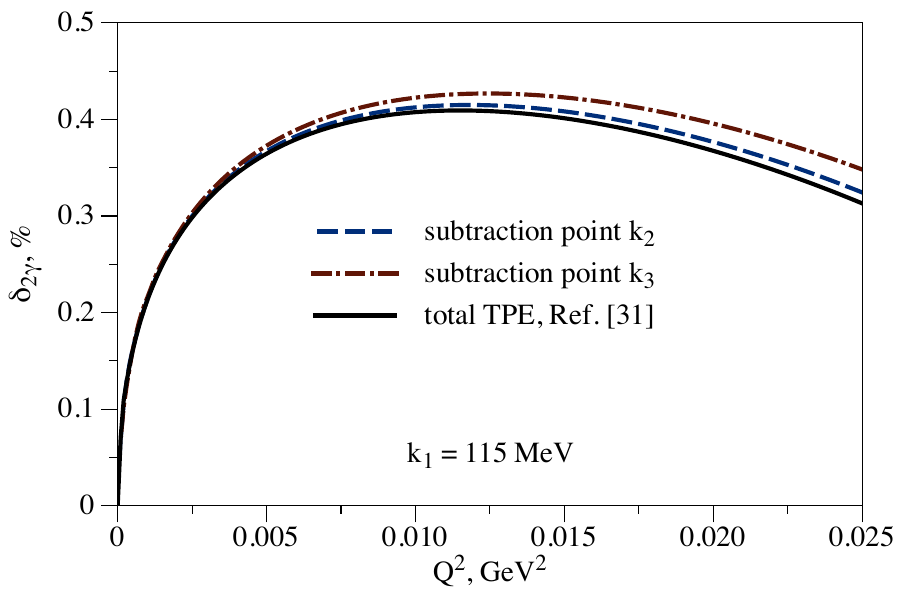}
\end{center}
\caption{The TPE correction to the unpolarized elastic $ \mu^{-}p $ cross section within the subtracted DR framework is compared with the model result of Ref. \cite{Tomalak:2015hva} for the MUSE beam momentum $ k_1 = 115~\mathrm{MeV}$.}
\label{absolute_xsection}
\end{figure}

In Fig. \ref{absolute_xsection}, we also present the absolute value of the TPE correction for the beam energy $k_1=115~\mathrm{MeV}$, comparing the model calculation with the subtracted DR predictions. We perform our analysis for the expected kinematical region of MUSE experiment and end our curves respectively.

\section{Conclusions and outlook}
\label{sec6}

In this work, we have extended the fixed-$Q^2$ dispersion relation formalism to the case of elastic muon-proton scattering at low energies and evaluated the two-photon exchange amplitudes within this approach. We accounted for the leading elastic intermediate state. The imaginary parts of TPE amplitudes were reconstructed from the input of one-photon exchange amplitudes by means of unitarity relations.

Using the dipole form for the proton elastic form factors, the real parts were evaluated performing dispersive integrals. According to our analysis of the unitarity constraints, the helicity-flip amplitude $\cF^{2 \gamma}_4$ can be constant at infinity. Consequently, this amplitude requires a once-subtracted dispersion relation. Unsubtracted DRs give us all other relevant amplitudes. We have related the subtraction function to the known value of TPE at some lepton beam energy corresponding to MUSE setup and predicted the TPE correction for the other planned energies. The resulting TPE contribution in MUSE kinematics is in reasonable agreement with a previous model estimate of total TPE in Ref. \cite{Tomalak:2015hva}, which was used to fix the subtraction function. The developed subtracted DR formalism can be used in the analysis of experimental data exploiting the forthcoming measurement of TPE by the MUSE collaboration as a subtraction point. Due to the small contribution of inelastic excitations, the subtracted DR prediction is in good agreement with the near-forward estimate of the sum of elastic and inelastic intermediate state contributions \cite{Tomalak:2015hva} within 10\%.

\section*{Acknowledgments}

This work was supported by the Deutsche Forschungsgemeinschaft DFG in part through the Collaborative Research Center [The Low-Energy Frontier of the Standard Model (SFB 1044)], and in part through the Cluster of Excellence [Precision Physics, Fundamental Interactions and Structure of Matter (PRISMA)].

\appendix

\section{Photon-polarization density matrix}
\label{app:photon_polarization}

In this Appendix, we provide the photon-polarization density matrix in the laboratory frame and discuss the physical meaning of the parameters $\varepsilon$ and $\varepsilon_\mathrm{T}$, see Eqs. (\ref{epsilon_def}) and (\ref{transverse_epsilon}).

The unpolarized lepton-proton scattering cross-section of Eq. (\ref{OPE_xsection0}) is proportional to the product of the leptonic $L^{\mu \nu}$ and hadronic $H^{\mu \nu}$ tensors as
\ber
\sum \limits_{\mathrm{spin}} | T^{1 \gamma}|^2 = \frac{e^4}{Q^2} L_{\mu \nu} H^{\mu \nu},
\eer
with the hadronic tensor \cite{Heller:2018ypa}:
\ber
H^{\mu \nu} & = &  \frac{1}{2} \sum \limits_{\lambda,\lambda'}  \bar{N}(p^\prime,\lambda^\prime) J_{p}^\mu (p', p) N(p,\lambda) \nonumber \\
& \times & \bar{N}(p,\lambda) J_{p}^\nu (p', p)  N(p^\prime,\lambda^\prime),  \\
H^{\mu \nu} & = & - \left(g^{\mu\nu}-\frac{q^{\mu}q^{\nu}}{q^2} \right) 4M^2 \tau_P G_M^2 \nonumber \\
 &+& \frac{4 P^\mu P^\nu}{1+\tau_P}\left( G_E^2 +\tau_P G_M^2\right),
\eer
and the leptonic tensor:
\ber
L^{\mu \nu} & = & \frac{1}{2Q^2} \sum \limits_{h,h'}  \bar{u}(k^\prime,h^\prime) \gamma^\mu u(k,h)  \bar{u}(k,h) \gamma^\nu u(k',h'), \nonumber \\  \\
L^{\mu \nu} & = & - g^{\mu \nu} + 2 \frac{k^\mu k'^\nu + k'^\mu k^\nu }{Q^2}.
\eer

Following Refs.~\cite{Dombey:1969wk,Akhiezer:1974em}, we generalize the photon-polarization density matrix in the laboratory frame to the case of massive lepton:
\ber
\rho_{ij} & \equiv &  \left( 1 - \left( 1- \frac{1}{\tau_P} \right) \delta_{i 3} \right) \left( 1-  \left( 1 - \frac{1}{\tau_P} \right) \delta_{j 3}  \right) \nonumber \\
& \times & \frac{ 1 - \varepsilon_\mathrm{T}}{2} L_{ij},
\eer
with the transverse linear polarization parameter $ \varepsilon_\mathrm{T}$ of Eq. (\ref{transverse_epsilon}), which varies between 0 for the backward scattering and 1 for the forward scattering.
 
The spatial components of the photon-polarization density matrix $\rho_{ij}$, where we align the z-axis along the virtual photon momentum, are given by
\ber
 \rho = \begin{bmatrix}
    \frac{ 1 + \varepsilon_\mathrm{T} }{2} & 0 & \sqrt{ \frac{ \varepsilon_\mathrm{T} \left( 1 +  \varepsilon_\mathrm{T} \right)}{2 \tau_P}} F  \\
    0 & \frac{ 1 - \varepsilon_\mathrm{T} }{2}  & 0 \\
   \sqrt{ \frac{ \varepsilon_\mathrm{T} \left( 1 +  \varepsilon_\mathrm{T} \right)}{2 \tau_P}}  F& 0 & \frac{\varepsilon }{\tau_P} \\
  \end{bmatrix} ,
 \eer
 with $ \varepsilon $ of Eq. (\ref{epsilon_def}) and $F$ of Eq. (\ref{factor_f}).

The relative flux of longitudinal to transverse photons is expressed in terms of the density matrix elements as
\ber
\frac{\Phi_L}{\Phi_T} = \frac{\rho_{33}}{\rho_{11}+\rho_{22}} = \frac{\varepsilon}{\tau_P}.
\eer
It is described by the photon polarization parameter of Eq. (\ref{epsilon_def}).

In the following, we provide the physical interpretation considering the proton current of Eq. (\ref{proton_current}), which is given by
\ber \label{proton_current1} 
J_{p}^\mu  =   \bar{N}(p^\prime,\lambda^\prime) \left(  G_M  \gamma^\mu - F_2  \frac{P^{\mu}}{ M} \right) N(p,\lambda).
\eer

We introduce an orthogonal to $P^\mu$ and $q^\mu$ vector $n^\mu$: $\left( P \cdot n \right) =0 $ and $\left( q \cdot n \right) =0$. Contracting the proton current with four-vectors $q^\mu,~P^\mu$ and $n^\mu$, we obtain:
\ber
J_{p}^\mu q_\mu &=& 0, \\
J_{p}^\mu P_\mu &=& G_E M \bar{N}(p^\prime,\lambda^\prime) N(p,\lambda), \\
J_{p}^\mu n_\mu &=& G_M  \bar{N}(p^\prime,\lambda^\prime) \hat{n} N(p,\lambda) .
\eer

Consequently, in any rest frame with collinear initial and final proton momenta $ \vec{p}~||~\vec{p}~'$(e.g., the laboratory frame, the c.m.f. or the Breit frame), longitudinal photons couple to $G_E$ and transverse photons couple to $G_M$ only. The relative flux can therefore be read off from Eq. (\ref{OPE_xsection}) for the unpolarized cross section as
\ber
\frac{\Phi_L}{\Phi_T} = \frac{\varepsilon}{\tau_P}.
\eer

\section{TPE amplitudes in forward limit}
\label{app:forward_relations}

In this Appendix, we study the forward limit of the invariant amplitudes beyond the OPE approximation, and contributions with $ Q^2 = 0 $ poles, exploiting the helicity amplitudes expressions through the invariant amplitudes of Eqs. (\ref{hamp}). The $ Q^2 $-expansion of coefficients allows us to obtain the following "kinematically" leading terms for the helicity amplitudes:
\ber
\frac{T_1 + T_3}{e^2} & \to &  \frac{8\nu}{Q^2} \left(\cG_2 + \frac{m^2}{\nu} \cG_4 \right),\label{t1_low_q2}  \\
\frac{T_2}{e^2} & \to & \frac{4 M}{Q} \frac{ m^2 + \nu  }{\sqrt{\Sigma_s} } \cG_1 \nonumber \\ 
&+&  \frac{4 M}{Q} \frac{ M^2 + \nu }{\sqrt{\Sigma_s} }  \left( \frac{m^2}{M^2} \cF_4 - \frac{\nu}{M^2} \cF_2 \right),\label{t2_low_q2} \\
\frac{T_4}{e^2} & \to & - \frac{4m}{Q} \frac{ \left( M^2+ \nu \right)  \cG_2 + \left( m^2 + \nu \right) \cG_4 }{\sqrt{\Sigma_s} }.
\label{t4_low_q2} 
\eer
The conservation of the total angular momentum, i.e., $T_2,~T_4(Q^2 \to 0) \to 0 $, implies:
\ber
&& \cG_1 + \frac{ M^2 + \nu}{m^2 + \nu} \left( \frac{m^2}{M^2} \cF_4 - \frac{\nu}{M^2} \cF_2 \right)  = 0, \\
&& \cG_2 + \frac{ m^2 + \nu}{M^2 + \nu}  \cG_4 = 0.
\eer 
Assuming the absence of the kinematical $Q^2$-singularity for the helicity amplitude $T_1 + T_3$, we obtain:
\ber
&&  \cG_2 + \frac{m^2}{\nu} \cG_4 = 0.
\eer 
These equations and the conservation of the total angular momentum, i.e., $T_6~(Q^2 \to 0) \to 0 $, allow us to write down four model-independent relations of Eqs. (\ref{g1_zero_at_low_q2}-\ref{f6_f4_f3_zero_at_low_q2}) for the lepton-proton scattering amplitudes in the forward limit beyond the contributions with $ Q^2 = 0 $ pole.

The following relations between amplitude pairs are valid:
\ber 
 \nu \cF_2 \left( \nu, Q^2 = 0 \right)    -  m^2 \cF_4 \left( \nu, Q^2 = 0 \right) & = & 0, \label{f4+g2_zero_at_low_q2_app} \\ 
 M^2 \cF_2 \left( \nu, Q^2 = 0 \right)  + m^2 \cF_5 \left( \nu, Q^2 = 0 \right) & = & 0.\label{f2+f5_zero_at_low_q2_app}
\eer

It is instructive to relate six non-forward lepton-proton amplitudes in the forward limit to three forward scattering amplitudes \cite{Tomalak:2017owk}. $ \cG^{2 \gamma}_M \left( \nu, Q^2 = 0 \right)$ and $ \cF^{2 \gamma}_6 \left( \nu, Q^2 = 0 \right)$ in the forward limit are directly related to the forward helicity-flip amplitudes $ f^{2 \gamma}_-,~g^{2 \gamma} $ of Ref. \cite{Tomalak:2017owk}, which are expressed as integrals over the proton spin structure functions, by
\ber
\cG^{2 \gamma}_M \left( \nu, Q^2 = 0\right) & = & \frac{f^{2 \gamma}_{-} (\omega)}{e^2}, \label{fminus_from_gm} \\
\cF^{2 \gamma}_6 \left( \nu, Q^2 = 0\right) & = & - \frac{M}{m} \frac{g^{2 \gamma} (\omega)}{e^2}. \label{g_from_f6}
\eer
Consequently, accounting for the four low-$Q^2$ constraints of Eqs. (\ref{g1_zero_at_low_q2}-\ref{f6_f4_f3_zero_at_low_q2}) all TPE amplitudes in the forward limit can be reconstructed from the experimental data on the proton spin structure. 

In order to obtain the spin-independent forward amplitude $ f^{2 \gamma}_+$ \cite{Tomalak:2017owk}, one should subtract the leading singular in $ Q $ behavior coming from the scattering of two point-like charges, see Ref. \cite{Tomalak:2015hva} for exact expressions. \footnote{The TPE contribution to the unpolarized helicity amplitude has no simple limit and scales as $ \left( T_1 + T_3 \right)^{2 \gamma} \sim Q^{-1}$ at low $Q$ due to the leading contribution from the scattering of two point charges.} The formal limit is given by
\ber \label{lamb_shift_and_TPE}
f^{2 \gamma}_+ \left( \omega \right)  &\equiv & \frac{T_1 + T_3}{2} \left( \nu,~Q^2 \to 0\right)\to  4 \nu e^2 \frac{ \cG^{2 \gamma}_2 + \frac{m^2}{\nu}  \cG^{2 \gamma}_4}{Q^2} \nonumber \\
&=& 2 \nu  e^2 \left.  \frac{\delta_{2 \gamma}(\nu, Q^2)}{ Q^2} \right \vert_{Q^2 \to 0},
\eer
where the second term is the $Q^2 \to 0$  behavior of $ \delta_{2 \gamma}$, see Eqs. (\ref{delta_TPE_massive}) and (\ref{delta_TPE_massive_low_Q2}). The amplitude $f_+$ at threshold ($\omega = m$) determines the leading in $\alpha$  TPE contribution to the Lamb shift of S-energy levels. We checked, that the particular contribution of the subtraction function in the forward doubly virtual Compton scattering \cite{Carlson:2011zd} (as well as modelling it by the exchange of a $\sigma$-meson \cite{Borisyuk:2017sda}) is obtained from the TPE correction to the unpolarized cross section \cite{Tomalak:2015hva} ($\sigma$-meson exchange contribution \cite{Koshchii:2016muj}) exploiting Eq. (\ref{lamb_shift_and_TPE}).

\section{TPE amplitudes in high-energy limit}
\label{app:high_energy_relations}

\begin{comment}
\ber \label{HE_invariants}
e^2 \cG_M & = & \frac{1}{2} \left( T_1 - T_3 \right), \\
e^2 \cF_2 & \to & - \frac{M}{2 \nu} \left( Q T_2 + m \left( T_5 - T_6 \right) \right) + \frac{M^2 Q^2}{4\nu^2} T_3,  \\
e^2 \cF_3 & \to &  \frac{M^2}{2 \nu} \left( T_3 - T_1 \right)   + \frac{M^2 Q^2}{8 \nu^2} \left( T_1 + T_3 \right),   \\
e^2  \cF_4 & \to &  \frac{M}{2 m}  \left( T_6 - T_5 \right) - \frac{M^2}{2 \nu} \left( T_1 - T_3 \right) - \frac{M Q}{2\nu} \left( T_2 - \frac{M}{m} T_4 \right),  \nonumber \\ 
e^2  \cF_5 & \to &  - \frac{M^2 }{2 m \nu}  \left( Q T_4 - M \left( T_5 - T_6 \right) +\frac{ m Q^2}{2 \nu} T_3 \right),  \\
e^2  \cF_6 & = & - \frac{M}{2m} ( T_5 + T_6 ),  \\
e^2 \cG_1 & \to & \frac{Q^2}{8 \nu}  \left( T_1 + T_3 \right),  \\
e^2 \cG_2 & \to &  \frac{Q^2}{8 \nu}  \left( T_1 + T_3 \right),  \\
e^2 \cG_3 & \to &  \frac{Q^2}{8 \nu}  \left( T_1 + T_3 \right) - \frac{1}{2} \left( T_1 - T_3 \right), \\
e^2 \cG_4 & \to &- \frac{Q  \left( 4 M T_4 + Q \left( T_5 - T_6 \right) +  \frac{M m Q}{\nu} \left( T_1 + T_3 \right) \right)}{8 M m \left( 1 + \tau_P \right)}. \nonumber \\
\eer
\end{comment}

In this Appendix, we study the high-energy limit, corresponding to $ \nu \to \infty $, of the invariant amplitudes beyond the OPE approximation, exploiting the invariant amplitudes definitions of Eqs. (\ref{stramp}, \ref{amplitudes_G1}-\ref{amplitudes_G4}). The leading terms in the high-energy expansion are given by
\ber \label{HE_invariants}
e^2 \cG_M & = & \frac{T_1 - T_3}{2}, \\
e^2 \cF_2 & \to & - \frac{M Q}{2 \nu} T_2 - \frac{m^2}{\nu}\frac{M}{m} \frac{T_5 - T_6}{2} \nonumber \\
&+& \frac{ M^2 m^2}{ \nu^2} \frac{Q}{2m} T_4+ \frac{M^2 Q^2}{4 \nu^2} T_3,   \\
e^2 \frac{\nu \cF_3}{M^2}  & \to & - \frac{T_1 - T_3}{2} + \frac{Q^2}{4 \nu} \frac{T_1 + T_3}{2} - \frac{M Q}{2 \nu} T_2 \nonumber \\
&-&  \frac{m^2}{\nu} \frac{M}{m} \frac{T_5 - T_6}{2} + \frac{m^2}{\nu}\frac{Q}{2 m} T_4 ,  \\
e^2  \cF_4 & \to & - \frac{M Q}{2 \nu} T_2  - \frac{M}{m} \frac{T_5 - T_6}{2}  + \frac{M^2}{\nu} \frac{Q}{2 m}T_4,   \\
e^2 \frac{ \nu  \cF_5}{M^2} & \to & \frac{M Q}{2 \nu} T_2 + \frac{M}{m} \frac{T_5 - T_6}{2} - \frac{Q}{2 m} T_4-  \frac{Q^2}{4\nu} T_3 , \nonumber \\  \\
e^2  \cF_6 & = & - \frac{M}{m} \frac{T_5 + T_6}{2} ,\\
e^2 \cG_1 & \to & \frac{Q^2}{4 \nu} \frac{T_1 + T_3}{2}- \frac{M Q}{2 \nu} T_2-  \tau_P \frac{ M^2 m^2}{ \nu^2} \frac{Q}{2m}  T_4, \nonumber \\ \\
e^2 \cG_2 & \to &  \frac{Q^2}{4 \nu} \frac{T_1 + T_3}{2} + \tau_P \frac{M Q}{2 \nu} T_2\nonumber \\
&+&\tau_P \frac{m^2}{ \nu} \frac{M}{m} \frac{T_5 - T_6}{2}  + \frac{m^2}{\nu} \frac{Q}{2 m}T_4,  \\
e^2 \cG_3 & \to & - \frac{T_1 - T_3}{2} +  \frac{Q^2}{4 \nu} \frac{T_1 + T_3}{2} \nonumber \\
&-& \frac{M Q}{2 \nu} T_2-  \tau_P \frac{ M^2 m^2}{ \nu^2} \frac{Q}{2m} T_4, \\
e^2 \cG_4 & \to &- \frac{  \frac{Q^2}{4 \nu} \frac{T_1 + T_3}{2} +  \tau_P \frac{M Q}{2\nu} T_2 }{1+ \tau_P} \nonumber \\
 && -  \frac{ \tau_P  \frac{M}{m} \frac{T_5 - T_6}{2} + \frac{Q}{2 m}T_4}{1+ \tau_P} .
\eer
The unitarity bounds were investigated in detail for the short-range interaction and can be strictly applied to the hadronic line only. We assume them to be valid in our process. The forward unpolarized helicity amplitude $A$ is constrained by the unitarity condition: $A \lesssim \nu \ln^2 \nu$, which is known as a Froissart bound \cite{Froissart:1961ux}. \footnote{The generalization to the case of amplitudes at a fixed $Q^2$ was performed in Refs. \cite{Mahoux:1969um,Azimov:2011nk}. However, the choice between the latter and the Froissart bound does not change qualitatively the high-energy and dispersive analysis of invariant amplitudes.} 

The amplitude $\cF_6$ is determined by $T_5 + T_6$ and corresponds to the exchange of a pseudoscalar particle:

\ber
e^2 \cF_6 & = & - \frac{M}{m} \frac{T_5 + T_6}{2} < \ln^2 \nu,
\eer
which allows us to write down the unsubtracted DR for this amplitude. As the leading high-energy behavior in the Regge limit is expected from the pomeron, it corresponds with no flip of the proton helicity.

We therefore assume for the proton non-spin-flip amplitudes $T_1+T_3$ and $T_4$:
\ber
T_1+T_3,~T_4 \lesssim \nu \ln^2 \nu,
\eer
and that other amplitudes are suppressed by some power of $\nu$:
\ber
T_5-T_6,~T_2 &\lesssim& \nu^\psi \ln^2 \nu,  \\
T_1 -T_3 &\lesssim&\ln^2 \nu,
\eer
with $ \psi < 0$, which leads to the $ \psi$-independent constraints of Eqs. (\ref{g1_g2_HE}).

\section{Hadronic model vs dispersion relations}
\label{app:hm_vs_drs}

\begin{figure}[H]
\begin{center}%\centering
\includegraphics[width=0.45\textwidth]{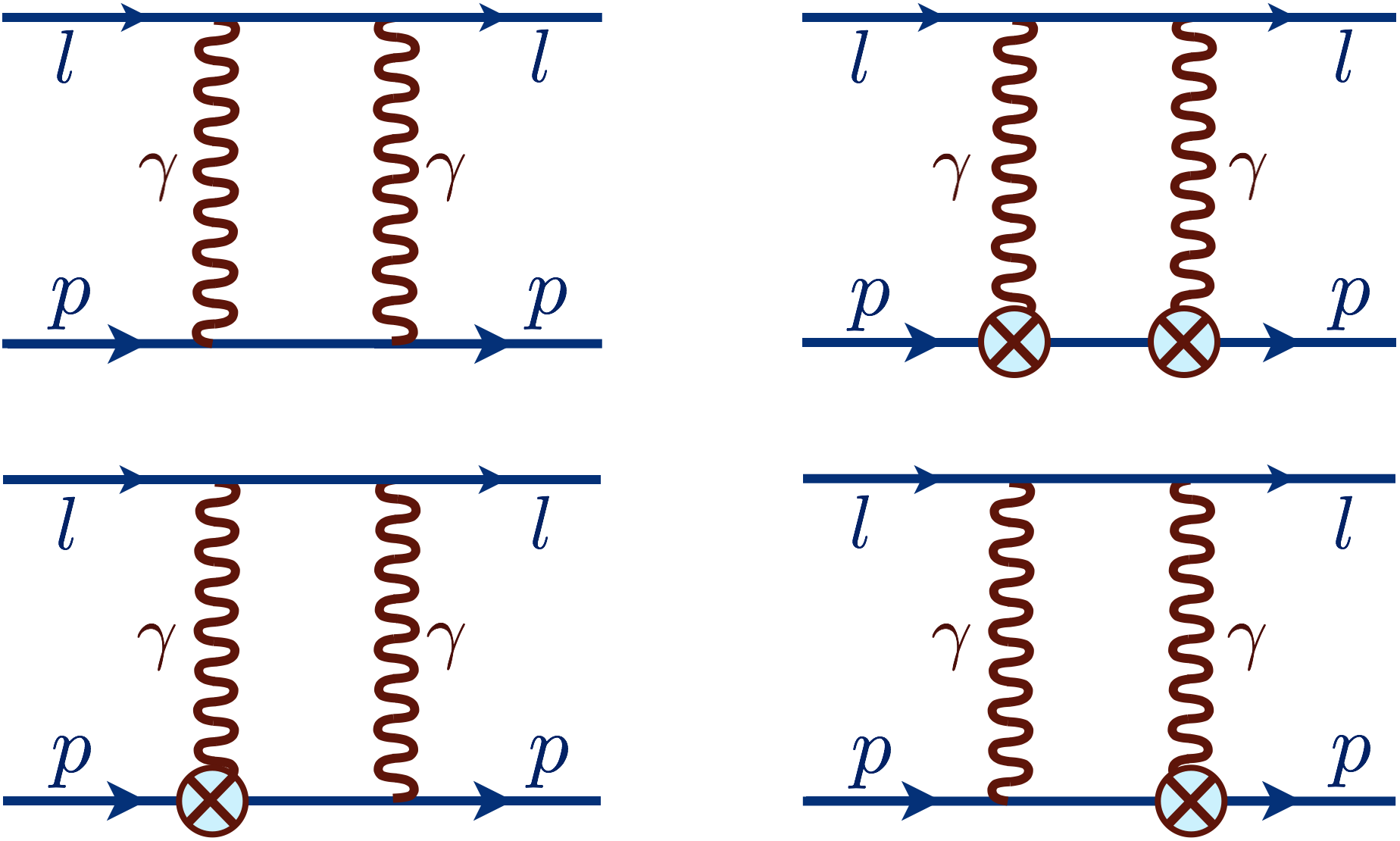}
\end{center}
\caption{Different contributions to the proton box diagram, depending on the different virtual photon-proton-proton vertices. The vertex with (without) the cross denotes the contribution proportional to the $ F_P $ ($F_D$) FF. The different diagrams show the $  \mathrm{F_D F_D} $ (upper left panel), $  \mathrm{F_P F_P} $ (upper right panel) and $  \mathrm{F_D F_P} $ (lower panels) vertex structures.}
\label{vertices}
\end{figure}

In this Appendix, we compare the unsubtracted dispersion relation approach to the hadronic model evaluation \cite{Blunden:2003sp,Tomalak:2014dja} of the proton intermediate state contribution to TPE amplitudes as well as to the unpolarized cross section. We study separately contributions whether $ F_D $ or $ F_P $ form factors, see Eq. (\ref{OPE_amplitude}), enter photon-proton-proton vertices. We denote the contribution with two vector couplings by $ \mathrm{F_D F_D} $, two tensor couplings by $  \mathrm{F_P F_P}  $, and the contributions from the mixed case by $  \mathrm{F_D F_P}  $, see Fig. \ref{vertices}.

\begin{figure}[H]
\begin{center}%\centering
\includegraphics[width=0.45\textwidth]{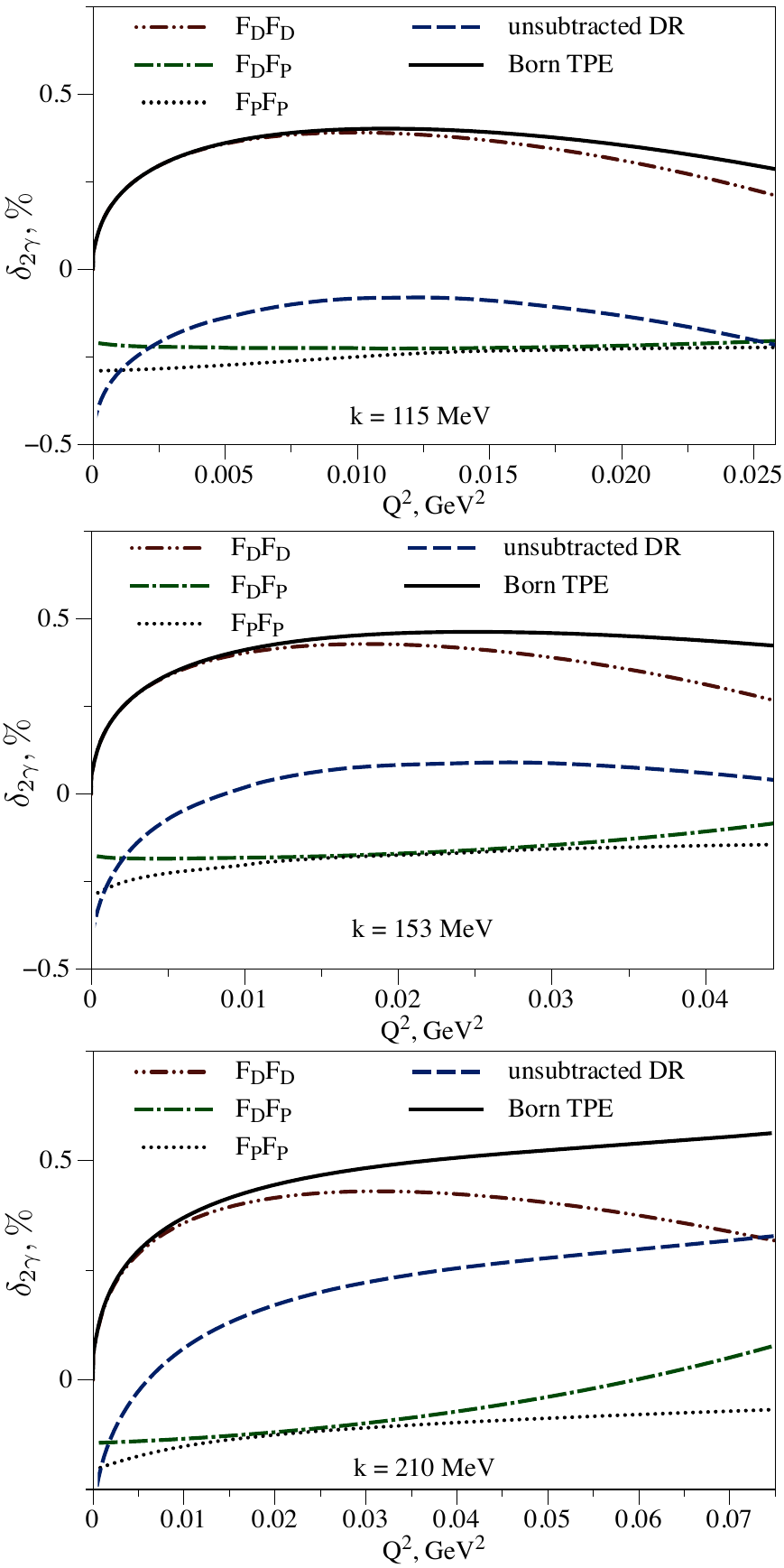}
\end{center}
\caption{TPE correction to the unpolarized elastic $ \mu^{-}p $ cross section evaluated for three nominal muon beam momenta within the unsubtracted DR framework. The total correction is shown by the blue dashed curves, the contribution from the $\mathrm{F_D F_D}$ structure of photon-proton-proton vertices is shown by the red double-dotted curves, the contribution from the $\mathrm{F_D F_P}$ structure by the green dashed-dotted curves, and the contribution from the $\mathrm{F_P F_P}$ structure by the black dotted curves. For comparison, the evaluation in the box diagram model (Born TPE) is shown by the solid black curves.}
\label{tpe_FFs_DRvertex}
\end{figure}

We provide the  unsubtracted DR prediction for $ \delta_{2 \gamma} $ in terms of different vertex structures in Fig. \ref{tpe_FFs_DRvertex} and compare it with the box graph model results \cite{Tomalak:2014dja}.

The contribution from the $\mathrm{F_D F_D}$ vertex structure in the unsubtracted DR formalism is the same as in the box graph model. In contrast to the model calculation, the negative contribution from the $\mathrm{F_P F_P}$ vertex structure cannot be neglected in the unsubtracted DR formalism due to the sizeable difference in $ \cF_4^{\mathrm{F_P F_P}}$ amplitude. The contribution from the $\mathrm{F_D F_P}$ vertex structure in the unsubtracted DR evaluation is negative as opposed to the box graph model. The amplitude $ \cF^{2 \gamma}_4$ is the main source of the difference between two approaches.

The resulting amplitudes in the box diagram calculation are in agreement with the low-momentum transfer limit of Eqs. (\ref{g1_zero_at_low_q2}-\ref{f6_f4_f3_zero_at_low_q2}) separately for contributions from $\mathrm{F_D F_D}$, $\mathrm{F_D F_P}$ and $\mathrm{F_P F_P}$ vertex structures.

Besides the model with dipole form factors of Eqs. (\ref{dipole_FFs}), we performed a calculation treating the proton as a point-like particle, e.g.:
\ber \label{point_model}
 G_E (Q^2) = 1, \quad G_M (Q^2)  = \mu_P.
\eer

Now, we compare the results for real and imaginary parts of amplitudes in case of $ \mathrm{F_D F_D} $, $ \mathrm{F_D F_P} $ and $ \mathrm{F_P F_P} $ vertex structures in the proton model with dipole and point-like FFs. For imaginary parts, the unitarity relation calculation and the box graph evaluation are in perfect agreement, since only on-shell information enters the evaluation by both methods. 

By comparing the DR results for real parts with the loop diagram evaluation for $ \mathrm{F_D F_D} $ vertex structure (for the sum of direct and crossed box diagrams), we found that all amplitudes are the same in both calculations. 

Also the real parts of the amplitudes $ \cG^{2 \gamma}_1,~\cG^{2 \gamma}_2,~\cG^{2 \gamma}_M,~\cF^{2 \gamma}_2,~\cF^{2 \gamma}_3,~\cF^{2 \gamma}_5 $ in the box graph model in case of $ \mathrm{F_D F_P} $ vertex structure are in agreement with the unsubtracted DR results. However, the unsbtracted DRs for the amplitudes  $  \cF^{2 \gamma}_4, ~\cF^{2 \gamma}_6 $ in the model with point-like proton do not converge (as well as for the $ \mathrm{F_P F_P} $ vertex structure), while the dispersive integral for the amplitude $\cF^{2 \gamma}_4 -\cF^{2 \gamma}_6 $ converges. Moreover, the result for real parts of the amplitudes $  \cF^{2 \gamma}_4, ~\cF^{2 \gamma}_6 $ in the box graph model with dipole form factors is shifted by a constant from the result of unsubtracted DR approach. 
Consequently, the real parts of the amplitudes $  \cF^{2 \gamma}_4, ~\cF^{2 \gamma}_6 $ in case of $ \mathrm{F_D F_P} $ vertex structure are in agreement between two types of evaluation if one uses the once-subtracted DR (the amplitudes $  \cF^{2 \gamma}_4, ~\cF^{2 \gamma}_6 $ are ultra-violet (UV) finite in the model with point-like proton). In case of $ \mathrm{F_P F_P} $ vertex structure, the unsubtracted DRs reproduce the box diagram model results for the amplitudes $ \cG^{2 \gamma}_1, ~\cG^{2 \gamma}_2, ~\cF^{2 \gamma}_2,~\cF^{2 \gamma}_5 $. While the results for real parts of the even amplitudes $ \cF^{2 \gamma}_4, ~\cF^{2 \gamma}_6 $ and $ \cF^{2 \gamma}_3 $ (the odd amplitude $\cG^{2 \gamma}_M$) are shifted by a constant (linear in $\nu$ function). The amplitudes $  \cG^{2 \gamma}_M, ~\cF^{2 \gamma}_3, ~\cF^{2 \gamma}_6 $ are UV divergent in case of the point-like proton with the following relations between the UV-divergent pieces: $ M^2 \cG^{\mathrm{UV}}_M = -\nu \cF^{\mathrm{UV}}_3 = \nu \cF^{\mathrm{UV}}_6$. Moreover, the amplitudes $ \Re \cG_M$ and $\Re \cF_3$ violate unitarity in case of $ \mathrm{F_P F_P} $ vertex structure in the model with point-like proton, while the unitarity constraints of Eqs. (\ref{g1_g2_HE}) are valid for  $ \mathrm{F_D F_D} $ and $ \mathrm{F_D F_P} $ vertex structures. Reconstruction of real parts within dispersion relations is consistent with unitarity constraints on the high-energy behavior.

All real parts of the TPE amplitudes in the box graph model are reconstructed using once-subtracted DRs. The three amplitudes $ \cG^{2 \gamma}_1,~\cG^{2 \gamma}_2,~\cF^{2 \gamma}_5 $ among the five TPE amplitudes, which are required for the evaluation of the cross-section correction by Eq. (\ref{delta_TPE_massive}), are reconstructed in the box graph model within the unsubtracted DRs. The DR analysis for the even amplitudes $ \cF^{2 \gamma}_3,~\cF^{2 \gamma}_4, ~\cF^{2 \gamma}_6 $ agrees with the box graph model only performing one subtraction. Fixing the subtraction constant to the hadronic model calculation, the Born TPE is reproduced by the subtracted DR analysis of Section \ref{sec:subtracted} up to $5 \times 10^{-4}$ relative accuracy level in the kinematics of MUSE. The remaining difference comes from the amplitude $ \cF^{2 \gamma}_3 $.

\end{document}